\documentclass[preprint]{aastex}
\usepackage{psfig}

\newcommand{\tfm}{\hbox{24$\micron$}$\,$}
\newcommand{\et}{\hbox{et al.}$\,$}
\newcommand{\etal}{\hbox{et al.}$\,$}
\newcommand{\Ha}{\hbox{H$\alpha$}$\,$}
\newcommand{\Hdn}{\hbox{H$\delta_{norm}$}$\,$}
\newcommand{\EWOII}{\hbox{EW([O II])}$\,$}
\newcommand{\EWHd}{\hbox{EW(H$\delta$)}$\,$}
\newcommand{\DHd}{\hbox{$\Delta\EWHd$}$\,$}
\newcommand{\kms}{\rm{\hbox{km s$^{-1}$}}}
\newcommand{\OII}{\hbox{[O II]}$\,$}
\newcommand{\Hd}{\hbox{H$\delta$}$\,$}

\shortauthors{Oemler et al.}
\shorttitle{Abell 851 and the Role of Starbursts}

\begin{document}

\title{Abell 851 and the Role of Starbursts in Cluster Galaxy Evolution} 
\author{Augustus Oemler Jr., Alan Dressler, Daniel Kelson, \& Jane Rigby}
\affil{The Observatories of the Carnegie Institution of 
Washington, 813 Santa Barbara St., Pasadena, California 91101-1292}
\email{oemler@ociw.edu, dressler@ociw.edu, kelson@ociw.edu, \& jrigby@ociw.edu}

\author{Bianca M.\ Poggianti \& Jacopo Fritz} 
\affil{Osservatorio Astronomico di Padova, vicolo 
dell'Osservatorio 5, 35122 Padova, Italy}
\email{bianca.poggianti@oapd.inaf.it \& jacopo.fritz@oapd.inaf.it}

\author{Glenn Morrison}
\affil{Institute for Astronomy, University of Hawaii, Honolulu, HI 96822}
\affil{ Canada-France-Hawaii Telescope, Kamuela, Hawaii 96743}
\email{morrison@cfht.hawaii.edu}

\author{Ian Smail}
\affil{Institute for Computational Cosmology, Durham University, South Rd, Durham DH1 3LE, UK}
\email{Ian.Smail@durham.ac.uk}
\clearpage
\clearpage

\begin{abstract}

We use extensive new observations of the very rich $z \sim 0.4$ cluster of galaxies A851 to examine the nature and origin of starburst galaxies in intermediate-redshift clusters. New {\em HST} observations, \tfm {\em Spitzer} photometry and ground-based spectroscopy cover most of a region of the cluster about 10\arcmin\  across, corresponding to a clustercentric radial distance of about 1.6 Mpc. This spatial coverage allows us to confirm the existence of a morphology-density relation within this cluster, and to identify several large, presumably infalling, subsystems. We confirm our previous conclusion that a very large fraction of the starforming galaxies in A851 have recently undergone starbursts. We argue that starbursts are mostly confined to two kinds of sites: infalling groups and the cluster center. At the cluster center it appears that infalling galaxies are undergoing major mergers, resulting in starbursts whose optical emission lines are completely buried beneath dust. The aftermath of this process appears to be proto-S0 galaxies devoid of star formation. In contrast, major mergers do not appear to be the cause of most of the starbursts in infalling groups, and fewer of these events result in the transformation of the galaxy into an S0. Some recent theoretical work provides possible explanations for these two distinct processes, but it is not clear whether they can operate with the very high efficiency needed to account for the very large starburst rate observed.

\end{abstract}


\section{Introduction}

Thirty years of challenging observational work has confirmed the existence of rapid recent evolution of the star formation rate of galaxies in rich clusters (Butcher \& Oemler 1978, BO), and has also provided support for BO's hypothesis that this evolution was caused by the transformation of spiral into S0 galaxies. However, despite this progress, no consensus has emerged concerning the nature of the process or processes driving the evolution of the spiral galaxy population. BO hypothesized that spirials faded after ram pressure stripping by the intracluster medium removed their gas supply. However, Dressler \& Gunn (1983) soon showed that many of the blue cluster galaxies were undergoing starbursts, suggesting the working of processes more violent than the mere fading of star formation. Dressler \& Gunn suggested that, rather than stripping the interstellar medium from a galaxy, the cluster ram pressure induced a starburst which consumed the galaxy's gas.  Since then, a number of other processes have been suggested, including starvation of a galaxy's star formation by the removal of the outer gas halos that are posited to surround spirals and replenish the disk gas by infall (Larson, Tinsley \& Caldwell 1980), tidal shocks on a galaxy, either due to the cluster core (e.g. Byrd \& Valtonen 1990, Henriksen \& Byrd 1996), to unvirialized subclusters (Gnedin 2003), or to other galaxies (Richstone \& Malmuth 1983, Icke 1985, Moore \etal (1996), and galaxy-galaxy mergers (Dressler \etal 1999, D99, van Dokkum \etal 1999, Struck 2006).

These processes may be categorized in several ways, including method of gas removal, star formation history, and dependance on environment. In stripping and starvation, gas content and star formation rate go hand-in-hand; as gas is swept from a galaxy or the external supply is shut off, there is a monotonic decrease in star formation rate. On the other hand, mergers, ram pressure induced star formation, and tidal encounters (including harassment, the cumulative effect of many weak encounters [Moore \etal 1996]) work at least in part by a temporary {\em increase} in star formation rate, i.e. by a starburst which can consume much or most of the gas. The processes also differ in their dependance on environment. All rely on interactions of a galaxy with its surroundings, but harassment, ram pressure induced star formation and stripping work only in the hot, dense environment of a rich cluster core, while mergers are expected to be {\em least} effective in such a hot environment.

One might suppose that distinguishing between these very different processes would be easy, but this has not been so.  A less-than-complete survey of 25 recent papers on the subject shows that 8 papers support starvation as the explanation of cluster-driven evolution, 4 papers support mergers, 3 support tidal encounters, 2 favor stripping, 2 favor ram pressure induced starbursts, 1 supports harassment, 3 favor a combination of mergers plus stripping, 1 favors stripping plus starvation and 1 favors stripping plus tidal encounters. Obviously, a consensus is not at hand.

In several previous papers (Poggianti \etal 1999, P99, Dressler \etal 2004, D04) we have presented evidence supporting the importance of starbursts in cluster galaxy evolution. (Definitions of ``starbursts'' vary. We shall mean by this term a significant temporary increase in the star formation rate of a galaxy above its long-term past average.) A companion paper to the present one (Dressler \etal 2009, hereafter D09) uses $24 \micron$ {\em Spitzer} observations of A851 to further elucidate the connection between the optical spectral properties of cluster galaxies and the nature of the ongoing starbursts. D09 show that  galaxies with strong Balmer absorption lines and \OII emission, which P99 and D04 have previously identified as starbursts, are indeed young starbursts buried under appreciable dust extinction, while galaxies with strong Balmer absorption lines but no \OII emission, which P99 and D04 identify as post-starburst galaxies, divide into two very different classes. The majority are true post-starbursts, but a significant minority are declining starbursts in which optical emission from the remaining OB stars is completely hidden beneath dust. Together, these papers demonstrate that starbursts are a widespread phenomenon in galaxies at earlier epochs, both within and outside of clusters. They also show that the post-starburst history of many cluster galaxies is very different than that of field galaxies: most field galaxies apparently resume normal star formation after a burst but many cluster galaxies do not. These facts suggest that a combination of starbursts plus an additional, cluster-specific process form at least one of the mechanisms responsible for the evolution of the star formation in cluster galaxies. If correct, more information on the environmental dependance of the phenomenon would clearly be useful for pinpointing the cluster-specific post-starburst process, as well as for elucidating the cause of the starbursts.

In this paper we present new {\em HST} imaging and ground-based spectroscopy of the cluster Abell 851, at a redshift $z = 0.41$. A851 is not a ``typical'' cluster; it is richer, has more substructure, and has a larger population of starbursting and disturbed galaxies than the average $z = 0.4$ cluster. However, there is every reason to think that the processes occurring at the time of observation in A851 typify those occurring--- at a somewhat lower rate--- in all clusters at that epoch. It therefore represents a particularly useful laboratory for studying the evolution of cluster galaxies.

The data presented here cover a much wider area than have most previous observations of intermediate redshift clusters, allowing us to probe further out into the cluster, beyond the cluster core and into the region where one might expect newly infalling galaxies and groups to predominate.  Thus they provide significant new information on the environmental dependance of the evolutionary processes, and the larger data set and the addition of {\em Spitzer} IR photometry also allows us to strengthen our previous conclusions about the origin and role of starbursts. We will demonstrate that a majority of the galaxies in A851 with on-going or recent star formation have undergone significant starbursts, and that many are  associated with unmistakable mergers. The most recent starburst and merger events are concentrated in several kinematic and/or spatial subcomponents of the cluster, some of which we will suggest are currently infalling subclusters.

The paper is organized as follows.  In \S2 we describe the data used in this paper, and in \S3 we discuss the structure of the cluster derived from spatial and velocity distributions.  In \S4 we revisit the morphology/density relation for A851.  In \S5 we discuss the analysis of the spectra of A851 members, correlate their properties with other galaxy and cluster parameters, and compare them with those of field galaxies at the same epoch.  In \S6 we examine the morphology and spatial distribution of the starbursting population, and in \S7 try to deduce from these the origin of the starbursts.

\section{The Data}

\subsection{Imaging Observations and Galaxy Morphology}

We make use of 4 sets of {\em HST} imaging observations of A851. Cycle 1 WFPC-1 imaging of the center of A851 is described in Dressler, Oemler, \& Butcher (1994). While these date have been largely superseded by later, post-refurbishing observations, they remain useful for the area of the PC frame gap in later WFPC-2 observations. Later Cycle 4
{\em HST} imaging of the central field and one outer field in Abell 851 were part of the "Morphs" study of 10 clusters (Smail et al. 1997, S97). For the present work we use two additional sets of Hubble Space Telescope observations. In a Cycle 6 GO program,  7 additional outer fields were observed with WFPC-2. Together with the original Morphs observations, these form a nearly complete (except for the missing area in the PC frames) square covering 8 arcminutes on a side, approximately 2.8 Mpc at the z = 0.405 redshift of Abell 851.\footnote{We adopt a standard $\Lambda$ cosmology with a Hubble constant $H_{o} = 70~\kms Mpc^{-1}$ throughout this paper.} Each Cycle 6 observation consisted of two orbits of exposure (4400 s total exposure time) with the F702W filter, providing the equivalent to a rest-frame band between Johnson $B$ and $V$.  Finally, in Cycle 16, 2 additional fields were observed with ACS, for 2235 sec each, using the F606W filter. These fields were chosen to cover most of the Northwest filament described in \S3. The stitched-together mosaic of all these data is shown in Figure 1. 

A photometric catalog reaching $R_{702} \sim 25$ magnitude was constructed for the central 8 arcminute fields using SExtractor (Bertin \& Arnouts 1996).  Consistent with our previous work on this cluster, a morphological sample of 844 objects complete to $R_{702} = 23.5$ was drawn from a list of 1541 objects that extends to $R_{702} \sim 24.5$.  Morphological classification was done by Dressler and checked for consistency with the earlier work by Oemler.  Following S97 (to which the reader is referred for details), we produced a catalog of Revised Hubble types, disturbance measures, and comments. In the two ACS fields, photometry has been taken from the work of Morrison (1999) and  morphological classifications have only been done for the spectroscopic objects.

\subsection{Spectroscopic Data}

Spectroscopic observations of galaxies in the outer fields were obtained with the COSMIC spectrograph (Kells et al. 1998) on the 200-inch Hale telescope at Palomar Observatory, as described in Dressler et al. (1999, D99).  These  new multislit observations were made during 6 observing runs in 1999 December, 2001 January, 2001 April, 2002 February, 2003 January, and 2003 February, with accumulated exposure times of 10,000--15,000\,s per mask in seeing conditions that were typically $\sim 1.5$ arcsec FWHM.

These spectra have spectral resolutions of 12\,\AA\ FWHM and typical signal-to-noise ratios of $\sim$10 per resolution element in the continuum at $\lambda \sim$ 6000\,\AA. This resolution and S/N is more than adequate for measurements of redshift and the equivalent width of \OII emission, but just adequate for measurements of the \Hd\ absorption line.

Our new data add 101 spectra with measurable redshifts and spectral features in the field containing A851. Only 10 of the newly-targeted galaxies failed to yield a redshift (though several of the derived redshifts are of marginal reliability); 13 objects turned out to be Galactic stars. The selection of galaxies in all but the new outer fields is described in detail in Section 2.1 of P99. Briefly, there are mild biases in favor of late--type spirals in the central regions of A851, but they have little effect on the overall sample. The new observations of the outer fields we of galaxies drawn from a magnitude limited sample, and are unbiased by any other properties of the galaxies.

We have measured linestrengths of $\OII\lambda\lambda$3727 and \Hd using the {\bf viewspectra} program which is part of the multiobject spectral reduction package COSMOS \footnote{http://www.ociw.edu/Code/cosmos}.  This is a semi-automated procedure in which emission and absorption lines are fit with a gaussian over a specified wavelength interval, and the continuum is fit to a straight line between two flanking continuum bands. The user is able to interact with the fit, correcting the continuum levels and fitting interval in cases where data problems make the automated result unreliable. The wavelength intervals for the \OII doublet consist of continuum bands of 3690--3710\,\AA\  and 3745--3775\,\AA, and a line region of 3717--3727\,\AA, and for \Hd, continuum bands of 4000--4040\,\AA\  and 4120--4165\,\AA, and a line region of 4081--4121\,\AA. Particularly in the case of \Hd, the choice of continuum is complicated by the numerous metal lines in this wavelength region, which make finding the true continuum very difficult, particularly for low-resolution, low signal-to-noise spectra such as ours. The regions chosen have been picked using stellar and galaxy spectra of a range of types to guide the choice of the most suitable intervals. These probably still provide underestimates of the true continuum levels, and thus underestimates of the \Hd line strength, but the results should be reasonably consistent across a range of galaxy types. Furthermore, tests have shown that linestrengths measured by this process are systematically identical to those measured by the old, interactive method used in the Morphs program (P99). The errors in these measurements vary considerably, depending on the signal-to-noise of the spectrum, but typical errors are 1\,angstrom for \Hd and several angstroms for \OII, except in the case of very high \OII equivalent widths which often occur in spectra with very weak, and thus poorly determined continua. The errors in these linewidths can be much larger. The lower limit for detection and measurement is, again, variable, but is typically 3.0\,angstroms for [OII] and 1.0\,angstrom for \Hd. The width of the \Hd line has been measured only for the stronger lines and is typically accurate only to 1--2\,anstroms. Note that, because of the interactive nature of the viewspectra programs, undetectably weak lines have not, in general, been measured. Thus, many objects have reported line strengths of exactly 0.00 \AA. The entire catalog of A851 spectra from our two studies is presented in Table 1. 

Based on the redshift histogram we show in Figure 2, we define cluster members to be all galaxies with redshifts $0.385 < z < 0.420$, and identify 44 new cluster members.  This excludes in particular two outliers at z = 0.373 that are probable members of the A851 {\it supercluster} but not the cluster itself.  For our total collection of 101 cluster members with spectrum quality, $Q \le 3$, we derive a mean redshift of 0.4050 and a rest-frame velocity dispersion of 1287 \kms.  Our new sample also produced 57 new field galaxies, for a total in this field of 106 with $Q \le 3$. (The outermost fields, nw1 and nw2 yielded only 9 cluster members compared to 38 field galaxies, which highlights the difficulty of studying outlying members of intermediate-redshift clusters.). In Table 2 we assemble data on the A851 cluster sample, containing all galaxies from Table 1 which meet the cluster membership criteria stated above and have spectrum qualities, Q, of 3 or better. We will want to compare the properties of A851 cluster members with those of field galaxies observed at the same epoch. To do this, we take galaxies that are not cluster members, in the redshift range $ 0.30 \le z \le 0.55$. To construct an adequate field sample we use galaxies from all 10 clusters used in D99. This sample is presented in Table 3.

There has been discussion in the literature about the representative nature of the spectroscopic samples of the Morphs studies.  We showed in D99 that our spectroscopic sample {\em in toto} follows our morphological sample in all cataloged parameters and is essentially a magnitude-limited sample. In D04, we showed that our A851 sample does have a modest bias towards late-type galaxies, but that this bias has only a minor effect on the mean spectral properties of the sample. We shall therefore assume, in the following discussion, that our data set is a fair sample of the cluster population.

\section{The Structure of Abell 851}

Many signs of substructure in A851 are apparent in Figure 1, with clumps and filaments of galaxies radiating from the complex cluster core.  A larger field surveyed by Kodama \et (2001) with Subaru's Suprime-Cam confirms a clumpy distribution for members of the cluster identified by photometric redshifts.  The strongest evidence, however, comes from the  XMM X-ray map by Schindler \et (1998) presented in Figure 3, which shows irregular contours and several distinct emission centers. Even at a distance of 1--2 Mpc from its core, Abell 851 appears to have galaxy concentrations with their own x-ray-emitting gas halos, suggesting that these subclusters have not yet crossed the cluster core.  We use the center of the southwest X-ray peak, which seems coincident with the center of galaxy distribution, to define a center of Abell 851 of $09^{h} 42^{m} 58.0^{s}$ and $46^\circ 59' 01''$ which we use in the analysis that follows.

The redshift histogram of Figure 2 is highly suggestive of substructure in A851, but a sample of 101 cluster member 
velocities is too small for a decisive test of substructure. However, by combining spatial and velocity information 
for the galaxies, using the Dressler \& Shectman (1988b) method, evidence for substructure is clear.  In the DS test,
the real cluster is compared to Monte Carlo versions in which the velocities have been randomly ``shuffled'';
the statistic is a sum of deviations of local velocity and velocity dispersion from global means.  Figure 4 compares 
the data for A851 with a typical example of a shuffled cluster.  The variance calculated for A851 with this test is 
4.87; a value this high occurs in only 15 out of 10,000 simulations. 

We conclude that the evidence for substantial substructure in the very
rich cluster A851 is very clear. Two significant lumps are apparent in
Figure 4, at $\Delta \alpha \sim -50''$, $\Delta \delta \sim 120''$, and
at $\Delta \alpha \sim -250''$, $\Delta \delta \sim 200''$. These can be
seen more clearly in Figure 5, which presents all three cuts in
$\Delta \alpha$, $\Delta \delta$, and z space. We can unambiguously
assign galaxies to each of the two groups, which we shall call the
North group and the Northwest group (the latter is actually more filament-like in appearance). The position of the North group on the sky coincides with the NNE extension of the X-Ray contours visible in Figure 3, suggesting that this group contains X-Ray emitting hot gas. The X-Ray source in the upper right corner of Figure 3, in the general location of the Northwest group, appears to be a point source, perhaps coincident with the k+a galaxy \#468, rather than extended emission due to hot gas.

 Membership in each group is noted
in the last column of Table 2. Dividing cluster galaxies into members of the core and the North and Northwest groups, we obtain values for the richness, mean redshift, and velocity dispersion of each which are presented in Table 4. Note that the relative richnesses of the three systems are very approximate, because the redshift samples are not complete. Both of the outlying groups are very cold. The North group has a velocity relative to the main cluster of $-2570\,\kms$ and has a projected location very near to the cluster core. The Northwest group has a velocity relative to the main cluster of $-900\,\kms$ and at a significantly greater projected distance. The positions and velocities are consistent with the North group infalling almost radially from behind the core cluster, and the Northwest group infalling at an angle to the line of sight from behind the core cluster.

\section{The Morphology-Density Relation in A851}

The additional data now available on A851 allow us to revisit the issue of the morphology--density relation in intermediate--redshift clusters. In
Dressler \et (1997, D97)  we used morphological data from 10 clusters to investigate the evolution of the morphology/density relationship found
for rich clusters at low-redshift (Dressler 1980). By comparing environments at the same physical density, we were able to show that elliptical
galaxies occur with the same frequency in these younger clusters, suggesting that ellipticals are long-time residents, as Butcher \& Oemler had
suggested and as is also supported by many other arguments based on stellar populations referenced in D99.  In contrast, the fraction of spiral
galaxies is higher at a given density, with a comparable drop in S0 fraction, at this earlier epoch, indicating a strong evolution in the population
of disk galaxies which we identify as a the principal cause of the Butcher-Oemler effect.

D97 found that these trends are quite strong in regular, concentrated clusters at $z \sim 0.5$, but considerably weaker in irregular intermediate-redshift
clusters such as A851. D97 was not able to offer a good explanation for this difference with low-redshift clusters, which show strong population gradients
with density in both regular and irregular clusters.  Our larger sample with its wider spatial extent, covering a greater range in local density than our previous sample, can be used to re-investigate the effect for A851.  In Figure 6 we compare the new morphology-density relation to that found for a
sample composed of regular and irregular clusters (D97, Figs 6 \& 8).  As was the case for the original Dressler (1980) study of the morphology-density
relation for low-redshift clusters, clusters are normally too sparsely populated to study the relation for individual clusters, all the more for D97 where
only one central HST field comprised the whole sample for each cluster.  For this reason it is not possible to compare A851 ``before and after" --- the
previous sample in A851 was just too small and had to be used in composite with 5 other irregular clusters to produce the morphology-density relation.
Likewise, the D97 relation for regular clusters at intermediate-redshift came from adding the data of 4 clusters.  However, the new A851 extended
sample is just large enough to be examined individually, and although still necessarily noisier than the composite, Figure 6 shows that it is a better
match to the regular cluster sample than the irregular cluster sample.  A clear trend of increasing elliptical-galaxy fraction, and a decreasing spiral galaxy
fraction, is seen with increasing local density.  In other words, with the extended sample, the morphology-density relation for the irregular cluster A851
is now compatible with the regular clusters, suggesting that the segregation of morphologies was already present even in this apparently unrelaxed cluster.

We believe that a plausible explanation can now be advanced for the apparent difference in the earlier paper.  Since the true correlation is almost
certainly with true space density and not projected (areal) density, it is important that the projected density track the true space density with only
modest dispersion.  Higher-redshift clusters suffer greater contamination by field galaxies than their low-z counterparts, as discussed by Koo (1988).
It is possible, then, that for the small fields of the previous study only the concentrated clusters provided a reliable relationship between apparent
and true space density.  For the irregular clusters more of the spread was likely due to projection effects which diluted the morphology/density relation.
By expanding the range of sampled densities, our new data should be less affected by contamination and, indeed, there is little or no difference between
the morphology/density relation of this irregular cluster and that of the regular clusters.  The further implication is that the morphology-density relation
might well be in place for all clusters, irregular and regular, at $z \sim 0.5$.  Additional studies such as this one and --- eventually --- full redshift coverage
in these fields to eliminate the field galaxies, will be able to verify if this explanation is correct.

\section{The Spectral Properties of Galaxies in A851}

\subsection{Using \Hd\ and \OII\ as Stellar Population Indicators}

The rationale for using the strength of the \OII\ and \Hd\ lines as measures of stellar populations in galaxies was described by Couch \& Sharples (1987), and more recently by P99 and D04. Briefly, when spectral coverage does not extend to the \Ha\ line, as in most observations of higher redshift galaxies, the \OII\ doublet provides the most convenient and reliable measure of the current star-formation rate. The \Hd\ absorption line, which arises primarily in intermediate-aged A stars, provides the best measure of star formation over the past billion years. D04 also demonstrated that the width of the Balmer absorption lines provided significant information about the intermediate age populations, wider lines indicating younger mean ages (via stark broadening in the early A stars.)

In populations of ``normal'' starforming galaxies, which have constant or monotonically decreasing rates of star formation, \EWHd\ has a well-defined dependance of \EWOII, rising as \OII\ grows stronger until, in galaxies with very high star-formation rates, it begins to fall again as the \Hd\ absorption line is filled in by \Hd\ emission. This is illustrated in Figure 7a and 7b. Figure 7a presents the \EWHd\ vs \EWOII\ distribution for galaxies in the Las Campanas Redshift Survey (LCRS Shectman \et 1996). These are field galaxies (that is to say the sample is dominated by isolated galaxies and those in groups, with a very small contribution from rich clusters) with a mean redshift of 0.1. D04 derives the mean relation between \EWHd\ and \EWOII\ for normal starforming galaxies, which we shall denote as \Hdn, using SDSS data analyzed by Goto (2003). This is shown as the central curve in Figure 7a. We define the quantity \DHd to be the difference between the value of \EWHd for a galaxy and the value predicted by the curve in Fig7a. The outer curves represent values of $\Hdn \pm 2.0$\, \AA, which the Goto data suggest are about the $2 \sigma$ limits of the distribution of \EWHd\ at a given \EWOII. The region between these curve should therefore contain about 95\% of all normal starforming galaxies. Figure 7b presents the indices for the sample of low-redshift cluster galaxies observed by Dressler \& Shectman (1988a, DS). These galaxies fill the same region of linestrength space as does the LCRS sample, the only difference being the crowding of the cluster galaxies towards low \EWOII\ strengths because of the morphology-density relation.

The only galaxies known to lie beyond this normal region are those which appear to be undergoing, or to have recently undergone, a strong starburst. Such objects can be selected by signs of a merger (e.g. Liu \& Kennicutt 1995, LK95) or by strong far-infrared flux (e.g. Poggianta \& Wu 2000). The Liu \& Kennicutt sample is presented in Figure 7c.  As noted by LK95, this sample is distinctly different than other nearby galaxies samples: The average \EWOII\ is much stronger, and many galaxies have strong \Hd\ lines. LK95 identify many of these objects as starbursts or post-starbursts, and present evidence from $H\alpha/H\beta$ ratios and from far-infrared photometry that many starbursts are buried behind high dust extinction. This conclusion is consistent with those of P99, Poggianti \& Wu (2000) and Poggianti, Bressan, \& Franceshini (2001), who demonstrate that the combination of weak [OII] emission, strong \Hd\ absorption and far infrared emission can only be obtained in regions undergoing a starburst, in which the youngest OB stars are more heavily obscured than the somewhat older A stars. All of these authors conclude that the true star formation rate in such galaxies can be much higher than what would be deduced from the strength of the optical emission lines. 

Furthermore, unless star formation is completely buried behind dust (q.v. D09 and \S5.2) spectra with no emission lines and strong \Hd\ arise only in galaxies in which star-formation has been rapidly truncated, with no remaining OB stars, but with a population of A stars which persists for many hundreds of millions of years. Such spectra with moderate \Hd\ equivalent widths, less than 4--5\,\AA, can be produced by the sudden truncation of what was previously ``normal'' star formation. However, P99 demonstrate that spectra of this type with \EWHd greater than 4--5\,\AA\_ can only be produced by the truncation of a starburst. In galaxy populations with objects of this type, the range of \EWHd\ always extends well above $4$\,\AA. Since it is likely that not all starbursts will be strong enough to produce very strong \Hd\ lines, and since it is certain that young post-starbursts with strong \Hd\ must steadily decay into older post-starbursts with weaker \Hd, it follows that every \Hd-strong post-starburst must be accompanied by some number of \Hd-weak objects which are also post-starbursts. It is, therefore, reasonable to assume that most if not all of the \Hd-enhanced, emission-free galaxies of all \Hd strengths are, in fact, post-starbursts rather than merely truncated normal galaxies.

Guided by the models and the behavior of real galaxies, D99, P99, and D09 subdivide the \Hd--\OII parameter space into regions which correspond to different histories of star formation. These regions, and the spectral classes which we assign to them are summarized in Table 5. k-type galaxies have no star formation and are dominated by old stars, e(c) objects contain``normal'' star formation, and e(a) objects have starbursts partially obscured by dust. The e(b) class, containing galaxies with very strong \OII\, can have two possible origins: starbursts or low-metallicity, high-excitation galaxies such as dwarf irregulars. Of the three e(b) galaxies in A851 with {\em Spitzer} \tfm observations, D09 found 1 out of 3 to be a genuine starburst. The k+a and a+k classes are post-starburst objects; D99 set a lower limit of $5$\,\AA\  for \OII emission in these types, because weaker lines cannot be reliably detected in our data; however this means that some ongoing starbursts with weak \OII will fall into this class.

\subsection{Properties of A851 and Intermediate-Redshift Field Galaxies}

In Figure 8  we present the \EWHd\--\EWOII\ distributions of the A851 cluster members and intermediate redshift field samples from Tables 2 and 3. We include all galaxies from these tables except those with uncertain measurements of \EWOII\ or \Hd.   As in Figure 7, we also include the $\Hdn \pm 2.0$ bounds of nearby normal galaxies.  Note also that, because of the semi-interactive method of line-fitting that we have used, values of 0 for \EWOII\ and particularly for \EWHd\ represent undetectable lines, which may have true equivalent widths of an angstrom or two. Therefore, in these plots, overlapping points- such as at 0,0 have been spread apart slightly to give a better idea of the numbers. Comparison of Figures 7 and 8 shows that both these samples have \EWOII\--\EWHd\ distributions inconsistent with normal star formation, but qualitatively similar to that seen in the LK starburst/merger sample, showing, as have been previously deduced by P99. that {\em starbursts are a very common mode of star formation at intermediate redshifts, both within and outside of clusters.}

That the e(a) galaxies in Figure 8 are, indeed,  similar kinds of objects as the low-redshift LK95 starbursting merger products is demonstrated in the accompanying paper by D09, which uses the spectra reported in this paper and \tfm {\em Spitzer} observations of A851 galaxies to show that (1) the A851 e(a) galaxies contain, as LK95 deduce for their sample, buried star formation with typical extinctions, at \OII, of a factor of 4, and (2) the observed star formation rates in the e(a)'s are a factor of 2 higher than their long-term average, demonstrating that they are, indeed, undergoing starbursts. Sato \& Martin (2006) have used a sample of emission line galaxies in A851 discovered from narrow band imaging to reach the same conclusions: the majority of emission line objects in A851 are e(a)'s, and these e(a)'s suffer from extinction at \OII of, typically, a factor of 4.

The second main conclusion of P99 is equally apparent from Figure 8: although almost all \Hd-strong galaxies in the field have ongoing star formation, as evidenced by \OII\ emission, many \Hd-strong galaxies in clusters do not. Several of the k+a/a+k galaxies in A851 do have weak \OII emission ($\EWOII > -5$\,\AA), and D09 demonstrate that about 1/3 of the emission-free k+a/a+k galaxies do, in fact, have significant star formation which is sufficiently buried behind dust to be undetectable at 3727\,\AA.  This result is consistent with the claim of Smail \etal (1999) that at least some of the a+k/k+a galaxies are buried starbursts rather than post-starburst objects. Although the width of the Balmer lines in these galaxies, shown in Fig. 9,  suggests that they are among the youngest starbursts in the cluster, modeling by D09 shows that their present high star-formation rates are already much lower than at the peak of their bursts. We shall refer to these objects as buried, decaying starbursts.

Nevertheless, 2/3 of the a+k/k+a galaxies are undetected at \OII and \tfm, consistent with their being genuine post-starburst objects. Figure 9 shows that these objects have significantly narrower \Hd linewidths than do either the e(a) galaxies or those a+k/k+a objects with detected \tfm flux, implying an older A star population, as would be expected if they were, in fact, post-starburst objects rather than buried starbursts sufficiently weak to be undetected at \tfm. We shall also present, in the next section, morphological evidence that these objects are qualitatively different than the objects containing buried starbursts. Such emission-free \Hd-strong post-starburst galaxies are rare not only in the intermediate redshift field, and the low redshift field (Zabludoff \etal 1996) but also in low redshift populations with numerous starbursts such as the Liu-Kennicutt sample. They  exist in some low-redshift clusters (Poggianti \etal 2004), but only among the dwarf galaxy population. As far as present evidence goes, luminous post-starburst galaxies are very rare everywhere except in higher redshift clusters. This unique feature of such clusters suggests that the starburst--post-starburst sequence is central to the evolution of cluster galaxies.

Since we are interested in the evolution of the starforming galaxies in clusters, it is useful to separate the galaxy populations into two groups: {\em passive} galaxies, like today's E's and S0's with no ongoing or recent star formation, and galaxies have or have recently had ongoing star formation, which we shall denote {\em active} galaxies. We shall include in the passive group all galaxies with the spectroscopic class of ``k'', as defined in Table 5; all others are denoted active. The fraction of active galaxies varies both with environment and with redshift; it ranges from a low of 13\% in the low redshift DS clusters to a high of 73 \% in the intermediate redshift field.

 We plot the cumulative distribution of \DHd\  among ``active'' galaxies in Figure 10, in which solid and dotted lines represent the low-redshift DS clusters and LCRS samples, the open circles the LK merger sample, the filled circles the A851 sample, and the stars represents the intermediate-redshift field sample. The contrasts in Figure 10 are striking; among local galaxy populations, only one composed entirely of mergers has a distribution similar to the active galaxies in the $z=0.4$ field, and even this population is not as dominated by \Hd-strong starbursting objects as is the active galaxy population in A851.  More than half of all active galaxies in A851 and about one third of active field galaxies at $z \sim 0.4$ have had disturbed histories of star formation, most, at least, involving starbursts. This rapid change in the mode of star formation since the epoch observed at $z = 0.4$ is at least as drastic as is the change in cluster galaxy star formation rates over the same interval, and has received much less attention.

In the analysis that follows, we shall refer to a number of spectroscopically-defined galaxy classes. These will include, in addition to the {\em active} and {\em passive} classes just defined:
{\em buried decaying starbursts} --- galaxies with  detected \tfm flux, no detected \OII flux, and $\EWHd \ge 3$\,\AA; {\em visible starbursts} --- galaxies with detected \OII and $\EWHd \ge 4$\,\AA; {\em post-starbursts} --- galaxies with no detected \tfm or \OII flux and $\EWHd \ge 3$\,\AA. In addition, we shall distinguish between {\em young starbursts} and {\em old starbursts} by the width of the \Hd line, wider lines indicating a younger population of A stars.

\section{Properties of the Starbursting Population}

If starbursts are very common in A851 (and, indeed, in the $z \sim 0.4$ field), what is their origin? It has become a commonplace that strong starbursts are the result of strong gravitational interactions between galaxies (e.g. Sanders \& Mirabel 1996), and there is a very large literature on all aspects of the phenomenon, including starburst signatures in mergers (e.g. LK95), merger signatures in starbursts and post-starbursts (e.g. Young \etal 2004, Schweizer 2005), and ULIRG's as mergers (e.g. Kim 2003), with only a very rare dissent (e.g. Bergvall, Laurikainen, \& Aalto, 2003). Most such work does not distinguish between strong tidal interactions and mergers, simply because the observational evidence often does not allow one to make such a distinction, but conventional wisdom expects that mergers should be much more effective than tidal encounters, particularly high-velocity encounters, and some numerical modeling supports this view (e.g. Mihos 2004).

Most evidence comes from observations of low-redshift field galaxies, but there is an increasing body of evidence that tidal/merger driven starbursts also occur in low-redshift clusters. Moss (2006) finds that the merger/interaction rate between galaxies is much enhanced among those infalling into clusters. Moss \& Whittle (2000) find that nuclear starbursts are more common in higher-density clusters, and are almost always associated with distortions that are probably due to mergers; and Sakai \etal (2002) find that starbursts in an infalling group of galaxies in A1367 were caused by tidal interactions or mergers. At higher redshifts, van Dokkum \etal (1999) detect a very high merger population in a cluster at $z = 0.83$, but do not look for the spectral signature of starbursts. P99 found that starbursting galaxies in the Morphs sample of $z \sim 0.4$ clusters were often merging or tidally interacting. Mid-infrared observations of galaxies in the rich $z \sim 0.39$ cluster Cl0024+1654 (Coia \etal (2005) finds many IR-luminous objects, most of which look like merging/interacting galaxies (see also Geach \etal 2006); and Marcillac \etal (2007) find a large number of strong, dusty starbursts in a rich cluster at $z \sim 0.83$.

Although gravitationally interacting galaxies are the most popular explanation for starbursts, it is not the only viable one. The suggestion by Dressler \& Gunn (1983) that the hydrodynamical shock caused by ram pressure from the intracluster medium on the interstellar medium of an infalling galaxy might induce starbursts has been recently revived by a number of authors. Poggianti \etal (2004) present evidence that starbursts in the Coma cluster were caused by gas shocks during a cluster merger, and Mercurio \etal (2004) detect ram pressure induced star formation in a group infalling into a $z \sim 0.2$ cluster. Gavazzi \etal (2003) claim that the same infalling group in A1367 that Sakai \etal (2002) thought were undergoing mergers, is in fact, being subjected to a hydrodynamical shock.

That distinguishing gravitational from hydrodynamical causes of starbursts is not easy, even for nearby objects, is shown by the disagreement about what is happening in A1367. Very high resolution X-ray observations can be useful, by identifying gas shocks that coincide with the locations of starbursting galaxies, but little adequate data exists at for higher redshift clusters. Nevertheless, one might hope that the combination of the kinematics, morphology, and positions of the galaxies in question would lend support to one or the other mechanism.

\subsection{Morphology of the Starburst and Post-starburst Galaxies}

Figure 11 presents images of all k+a and a+k galaxies with no \OII emission in A851 which lie within the post-refurbishment {\em HST} frames. These images, and those in subsequent mosaics, are between $4\arcsec$ and $8\arcsec$ across, equivalent to 20 --- 40 kpc, at the redshift of A851. The top group contains the {\em buried, decaying starbursts}. The second group contains the genuine {\em post-starbursts}, galaxies with neither \OII nor \tfm flux. The difference in appearance between these groups is quite striking. Those in the top group are, for the most part, clear-cut mergers, with shells, tidal arms, and multiple central light components. In contrast, the objects in the second group are all rather normal-looking early-type disk galaxies; consistent with previous observations that post-starburst objects appear to be mostly S0's or proto-S0's (D99). As promised in the previous section, the clear-cut morphological differences between a+k/k+a galaxies with and without \tfm flux seen in Fig. 11 is consistent with our previous conclusion that the former are buried starbursts, but the latter are true post-starburst galaxies, rather than merely weaker buried starbursts.

Figure 12 presents images of all e(a) and e(b) galaxies in A851 which lie within the post-refurbishment {\em HST} frames, and which have redshift qualities $Q \le 3$. These {\em visible starbursts} form a more heterogeneous collection of objects than the buried starbursts in the top group of Figure 11. Some, like objects \#319 and \#334, look very much like the mergers in Figure 11, but others, like objects \#462, \#404, and \#469 are at best slightly disturbed, almost normal late spirals. Objects like \#305 and \#317 are difficult to interpret; the may be edge-on equivalents to the very disturbed object \#291, or they may simple be analogues of much more normal objects like \#464.

 That some --- or even most --- mergers should have detectable \OII is hardly surprising: all of the LK objects do. It is therefore natural to lump objects like \#319 and \#334 in with the buried starbursts in Figure 11, as examples of merger-driven starbursts. Objects like \#367 are clearly very disturbed and may be merger-driven or may have other causes. The nature of the two e(b) galaxies is also ambiguous; they may be disturbed or may simply be irregulars. How to interpret more normal-looking objects like \#462 and \#464 is much less clear- there is scant evidence that they have been involved in a recent merger, or indeed have been subject to a recent strong perturbation of any kind. However, it is well to remember that strong interactions between galaxies, capable of transferring much gas from one object to another, are not always obvious from optical imaging. For example, the interactions within the M81 group which have caused a major starburst within M82 would be completely undetectable in our {\em HST} images.

\subsection{Spatial Distribution of the Starbursting Population}

Given the substructure within A851 shown by Figures 3 through 7, one may hope that the distribution of spectral and morphological types will provide further insight into the origin of the starbursting population. In the maps presented below we separate out galaxies which are members of the North and Northwest groups, and display them in boxes, offset from their true positions to eliminate confusion with galaxies in the main cluster core. As usual, we only include galaxies with redshift quality $Q \le 3$. In each map, the open circles represent the entire group of A851 galaxies from which the objects of a particular type are drawn (for example, \tfm-luminous buried starbursts are drawn from objects within the {\em Spitzer} survey area.)

Figure 13a presents the distribution of starburst galaxies: e(a) and e(b) galaxies, and those a+k and k+a galaxies with detected \OII or \tfm flux. The larger filled circles are those galaxies with $\sigma(\Hd) > 11$\,\AA, which should be the youngest starbursts. Starbursts can be found throughout the cluster and its neighboring groups, but the youngest starbursts clearly favor the cluster core and the two groups. In contrast, the distribution of post-starburst galaxies (a+k and k+a galaxies with no \tfm flux), shown in Figure 13b, in much broader, avoiding the cluster center and favoring the outlying groups and the outer parts of the central cluster. (Sato \& Martin (2006) find that their e(a)'s favor the outer groups, and their kinematics suggest that they are on their first pass infalling into the cluster.)

The distribution of distortion classes is presented in Figure 13c. As with the starbursts, these objects populate both groups and are centrally concentrated in the main cluster. The clear-cut mergers, given "M" types, are very concentrated towards the core of the main cluster. (The other distortion classes do not have distinct distributions, and are all shown with one symbol). Not surprisingly then,  since Figure 11 shows them to all be obvious mergers, the buried starbursts, with strong \Hd, detected \tfm flux and no \OII, have an equally centrally concentrated distribution, as shown in Figure 13d.

\section{Origin of the Starbursting Population}

Figure 13  shows that many spectral and morphological subsets of galaxies have very different distributions. We first consider galaxies within the main cluster, excluding members of the North and Northwest groups. Figure 14 presents the cumulative radial distribution of a number of galaxy populations in the main cluster. The contrasts are striking: three almost-identical sets of objects, buried, decaying starbursts, young starbursts, and clear mergers, are all very concentrated towards the cluster center. Older starbursts are more broadly distributed, similar to the overall distribution of active galaxies, and post-starbursts avoid the cluster center completely. This radial segregation could have one of two causes. One cause would be a difference in the orbits of the different sets of galaxies: those concentrated in the center could have orbits which confine them there, and the post-starbursts could have orbits which avoid the center. This explanation fails for two reasons. Firstly, to be confined to the cluster center, a set of galaxies must have a smaller velocity dispersion than the general population. However, within 100\arcsec--- $\sim500 kpc$--- of the cluster center the velocity dispersion of the buried, decaying starbursts, young starbursts, and M-class objects is virtually identical to that of all active galaxies ( $1044 \kms$ vs $1020 \kms$). Secondly, young starbursts must be the progenitors of old starbursts, and old starbursts of post-starburst, so these sets of objects should have common orbits.

The only viable alternative explanation for the radial segregation is that of timing: the buried, starbursting mergers must be triggered as they pass through the cluster center (presumably on quite radial orbits) and decay into older starbursts, then post-starbursts as they move away from the center. This model works quite well: the relative radial segregation of types is what one would expect, and the timing works. A galaxy at $z=0.405$ traveling across the line of sight at $1100\,\kms$ will move  100\arcsec--- $\sim500 kpc$--- in about 500 million years. For a starburst lifetime of a few hundred million years, initiated at the cluster center, one would expect to see starbursting objects within the observed distances from the center, and one would expect post-starbursts to populate a region from less than 100 arcsec to several hundred arcsec from the cluster center ($500 kpc--1 Mpc$).

Why should falling through the core of a rich cluster precipitate mergers among the infalling galaxies? For single infalling objects, merging with another passing galaxy is the last thing one would expect, given the very large relative velocities. However, if galaxies fall in as bound pairs or small groups, the perturbation to the orbits of the bound galaxies may be sufficient to drive some fraction into orbits which lead to mergers. Struck (2006) has argued that the gravitational pull of the cluster core on a small group falling through the core will cause the group to shrink, increasing  its density by an order of magnitude, and the merger rate among its galaxies by a factor of 100 during the time of its passage. The greatest challenge for any model may be that of efficiency; Table 7 shows that 71\% of the active galaxies in the inner core of A851 are starbursts, and 78\% of those are clear mergers. These are very large fractions, but a factor of 100 increase in merger rate may be enough to produce them.

Can the same mechanism explain the starbursts seen in the North and Northwest groups? The association of starbursts with infalling groups is a common theme in recent work. Owen \etal (2005), Mercurio \etal\ (2004), Sakai \etal (2002), Poggianti \etal (2004), Moss (2006), and Wang, Ulmer, \& Lavery (1997) all discuss examples of this process. Some of these invoke ram pressure as the cause of the starbursts, others mergers. Carlberg (2004) has modelled the mergers of galaxies in small groups, and showed that it increases, although slowly, with redshift. However, Bekki (1999) has modeled the {\em tidal} forces on galaxies in a small group infalling into a rich cluster and shown that they are sufficient to drive a large starburst, {\em without} the need for a galaxy-galaxy merger. 

We have, then, at least one mechanism, Struck's, for merger-induced starbursts in cluster cores, and one mechanism, that of Bekki, for starbursts without mergers in infalling groups. Figure 15 presents images of all starbursting galaxies within 75 arcsec of the cluster center (top group) and in the North and Northwest groups (bottom group). (Among the latter we include, within a box, the two objects with no \OII which are beyond the {\em Spitzer} survey area. Since we have no information on their IR flux, we cannot rule out one or both of these are buried starbursts, although their appearance is much more like the genuine post-starburst objects in Figure 11.) The two groups of objects are quite different in appearance. Almost all of the core starbursts show clear-cut signs of mergers: tails, shells, multiple central lumps. In contrast, while several of North and Northwest group objects  are quite peculiar, very few are as compelling examples of mergers as are the core galaxies. Several look like quite normal late-type spirals. Many of the latter show signs of disk HII regions, while the core galaxies have much smoother disks, suggesting that the starbursts might be confined to the nuclei--- consistent with the higher extinction seen in some of these.

Table 6 summarizes the active galaxy content of the main cluster, the North and Northwest groups, and of field galaxies between $0.30 \le z \le 0.50$, from Table 3. Active galaxies are all except k-type, and starbursts are those with $\DHd \ge 2.0$. Post-starbursts are defined as those with $\DHd \ge 2.0$ and with no detectable \OII or \tfm flux. The two k+a galaxies in the Northwest group which are outside the {\em Spitzer} survey area are assumed to be genuine post-starbursts on the basis of their appearance. M-types include only those classified as certain M, and do not include those in the M? class. There are several points to note in this table. The starburst rate is even higher in the infalling groups than in the main cluster- as the papers cited earlier would predict- but it is much lower in the field, as Figure 10 has already demonstrated. The ratio of post-starburst to starburst galaxies is much higher in the main cluster than in either the groups or the field, and the fraction of starbursts with clear merger appearance is large in the main cluster but negligibly small elsewhere. 

The information contained in Figure 15 and Table 6 may be telling us that  all or many of the starbursts outside of cluster cores represent a different phenomenon than the core mergers. As has been clear for some time, the aftermath of starbursts is different in clusters and the field: most field galaxies continue to form new stars after a burst, but many or most cluster galaxies do not. There are many ways of perturbing a galaxy into a temporary increase in its star formation rate; most are unlikely to lead to the total exhaustion of a galaxy's gas supply, but mergers may well do so, by expelling some gas and driving the remainder to the galaxy core where it is consumed in the burst (Struck \& Brown 2004), particularly in cluster cores where ram pressure stripping can aid the gas removal. The much greater prevalence of merger-driven starbursts in the core of at least A851, may explain the greater ratio of post-starburst to starburst galaxies there.

\section{Discussion}

We can summarize what we now know about A851 as follows:

\begin{enumerate}

\item 
A851 is a complex system of merging subclusters, many with their own hot X-Ray gas content.

\item
Despite this, A851 shows a well-established morphology-density relation, with early type galaxies concentrated in the densest lumps.

\item
In two infalling groups, an extremely high fraction of the starforming galaxies are undergoing starbursts. The starburst rate in the main cluster is significantly lower, except in the very center, but still much higher than in the field at the same epoch.

\item
The ratio of post-starburst objects to those currently undergoing starbursts is much higher in the main cluster than in either the infalling groups or the field.

\item
Most of the starbursts in the core of the main cluster are young, are buried behind very high dust extinction, and occur in galaxies which appear to be undergoing mergers . In contrast, these characteristics apply to a minority of the group and field starbursts.

\item
There is a strong radial gradient in the age of the starbursts in the main cluster. The youngest occupy the very center, with older starbursts and post-starbursts occurring at progressively greater cluster-centric distances.

\end{enumerate}

These facts suggest that there are at least two causes of starbursts. In the cluster center, starbursts are the result of galaxy-galaxy mergers, or more precisely {\em major} mergers in which the mass ratio between the two galaxies is less than a few, since only such events produce the morphological signatures seen in Fig.11. For reasons mentioned above, such major mergers are likely to be quite effective at removing a galaxy's gas, and therefore provide a simple explanation for the gas-free post-starburst objects found mostly in clusters. Furthermore, if Struck's (2005) arguments are to be believed, the cores of rich clusters provide an unexpectedly hospitable environment for such major mergers. Such a core-centered process explains the radial gradient in the ages of starbursts seen in A851.

There are two worries about this otherwise satisfying picture. Firstly, it is commonly believed that major mergers result in bulge-dominated galaxies, not disky objects like the post-starburst objects seen in Fig. 11. Secondly, Stuck's mechanism, at least, requires that the bursting galaxies be members of {\em groups}. Given the small number of objects involved at any one time, one would expect them to be members of one or two groups. If the group(s) is infalling perpendicular to the line of sight one would expect a small velocity dispersion; if infalling along the line of sight one would expect a small velocity dispersion and a large velocity offset from the main cluster. Neither expectation is met by the central starbursts, whose velocity distribution is very similar to that of the entire cluster.
These are serious problems. Nevertheless, the morphological evidence for major mergers is rather compelling, and the location of the bursts and the radial gradient in the ages of the bursting and post-bursting galaxies strongly points to a process sited in the cluster core.

The other primary site of starbursts appears to be infalling groups. The starbursting group members, while often peculiar, show much less compelling evidence for major mergers than do the core objects. Furthermore, group starbursts seems to produce gas-free post-starburst remnants much less frequently than do core starbursts. The absence of ram-pressure stripping in these low-density groups may be a partial explanation for this latter difference, but the available facts are consistent with a different mechanism driving starbursts in this environment. Either minor mergers, with smaller mass ratios, tidal encounters, or the cluster tides proposed by Bekki may provide a mechanism which is more consistent with the group starburst characteristics. Bekki's mechanism is particularly appealing because it can explain why the starburst rate in infalling groups is substantially higher than in the field- which is, after all, composed mostly of groups.

A major question is the extent to which the processes observed in A851 are typical of other intermediate-redshift clusters of galaxies. It cannot be ignored that a substantial fraction of the 25 recent papers mentioned in   \S1 find evidence for different processes at work. For example, the very extensive studies of $z \sim 0.4$ clusters by Moran \etal (2005, 2007) provide compelling evidence for a gradual fading of star formation in some cluster galaxies. A851 is, admittedly, a rather unusual cluster, with more substructure and more disturbed galaxies than most. Nevertheless, astrophysical processes should be universal. It is reasonable to expect that what is observed occurring in the core of A851, and in its infalling groups, has also occurred during some phases of the building of all rich clusters. Whether the average rate at which these processes occur in most clusters is sufficient to entirely explain the transformation of spiral to S0 galaxies which characterizes the recent evolution of cluster galaxies, or whether it is only one of a number of processes working in concert is a question which only further observations of other clusters can answer.

\acknowledgments

We thank Ray Lucas at STScI for his enthusiastic help which enabled the efficient gathering of these {\em HST} observations. We also thank our earlier collaborators Scott Trager, Richard Ellis, and Warrick Couch for their help. Dressler and Oemler acknowledge the support of NSF grant AST-0407343, and Smail acknowldeges support from the Royal Society. Optical imaging of A851 is based on observations obtained with MegaPrime/MegaCam, a joint project of CFHT and CEA/DAPNIA, at the Canada-France-Hawaii Telescope (CFHT) which is operated by the National Research Council (NRC) of Canada, the Institute National des Sciences de l'Univers of the Centre National de la Recherche Scientifique of France, and the University of Hawaii

{\it Facilities:} \facility{HST (WFPC)}, \facility{HST (WFPC2)}, \facility{HST (ACS)}, \facility{Spitzer (MIPS)}, \facility{Hale (COSMIC spectrograph)}, \facility{MMT (Megacam)}

%
%

%
%
\begin{deluxetable}{rccccrrrccr}
\tablewidth{0pt}
\tabletypesize{\small}
\tablenum{1}
\tablecaption{A851 Region Spectroscopic Sample}
\tablehead{\colhead{Object}	&	\colhead{$\alpha$}	&	\colhead{$\delta$}	&	 \colhead{z} & \colhead{Q} & \colhead{\OII} & \colhead{\Hd} & \colhead{$\sigma(\Hd)$} 	&	\colhead{class}	&	\colhead{r}	&	\colhead{I.D.}\\
\colhead{(1)} & \colhead{(2)} & \colhead{(3)} & \colhead{(4)} & \colhead{(5)} & \colhead{(6)} & \colhead{(7)} & \colhead{(8)} & \colhead{(9)} & \colhead{(10)} & \colhead{(11)}}
\startdata
72	\phm{00}	&	09:43:02.63	&	+46:56:42.4	&	0.6505	&	3	&	-65	\phm{0}	&	4.2	\phm{0}	&	7.0	\phm{0}	&	e(b)	&	21.51	&	dg92\_71	\\
73	\phm{00}	&	09:43:03.54	&	+46:57:06.4	&	0.3577	&	2	&	0	\phm{0}	&	0.0	\phm{0}	&	...	\phm{0}	&	k	&	21.38	&	dg92\_103	\\
74	\phm{00}	&	09:43:04.73	&	+46:58:15.7	&	0.4581	&	2	&	0	\phm{0}	&	0.0	\phm{0}	&	...	\phm{0}	&	k	&	20.42	&	dg92\_234	\\
75	\phm{00}	&	09:43:01.14	&	+46:58:31.2	&	0.2784	&	2	&	0	\phm{0}	&	0.0	\phm{0}	&	...	\phm{0}	&	k	&	19.57	&	dg92\_270	\\
76	\phm{00}	&	09:43:05.37	&	+46:59:05.3	&	0.2816	&	2	&	0	\phm{0}	&	0.0	\phm{0}	&	...	\phm{0}	&	k	&	20.44	&	dg92\_349	\\
77	\phm{00}	&	09:43:02.46	&	+46:59:23.1	&	0.5087	&	2	&	0	\phm{0}	&	2.3	\phm{0}	&	...	\phm{0}	&	k	&	21.51	&	dg92\_392	\\
78	\phm{00}	&	09:43:08.02	&	+46:59:23.6	&	0.2803	&	1	&	0	\phm{0}	&	0.0	\phm{0}	&	...	\phm{0}	&	k	&	19.53	&	dg92\_396	\\
79	\phm{00}	&	09:43:06.37	&	+46:59:24.1	&	0.4577	&	1	&	0	\phm{0}	&	0.0	\phm{0}	&	...	\phm{0}	&	k	&	20.42	&	dg92\_398	\\
\enddata
\tablecomments{Col.(1) Object--- galaxy index number; Cols. (2,3) $\alpha$, $\delta$--- Right ascension and declination (epoch 2000); Col. (4) z--- redshift, Col. (5) Q--- a spectrum/redshift quality indicator (1=excellent, 4=poor); Col. (6) \OII--- the rest frame equivalent widths of \OII in angstroms,  a blank entry indicates an uncertain measurement; Col. (7) \Hd--- the rest frame equivalent widths of \Hd, in angstroms,   a blank entry indicates an uncertain measurement; Col. (8) $\sigma(\Hd)$---  the rest frame width of the \Hd absorption line, only measured for stronger lines; Col. (9) class--- the spectral class, as defined in D99, and described in \S5;  Col. (10) r--- Gunn r magnitude; Col. (11) ID--- the spectrum id, as described in D99 . The new spectroscopic data are flagged as 0939\_3 ms\_1, ms\_2, ms\_3, nw1, and nw2.}
\end{deluxetable}

%
%
\begin{deluxetable}{cccrrrrcrccrrc}
\rotate
\tablewidth{0pt}
\tablenum{2}
\tabletypesize{\footnotesize}
\tablecaption{A851 Cluster Sample}
\tablehead{\colhead{Object} & \colhead{z} &	\colhead{Q} &	\colhead{\OII} &	\colhead{\Hd} 	&	\colhead{\DHd} &	\colhead{$\sigma(\Hd)$} &	\colhead{class} 	&	\colhead{F$_{24\micron}$} 	&	\colhead{morph} 	&	\colhead{int} 	&	\colhead{$\Delta\alpha$} 	&	\colhead{$\Delta\delta$} 	&	\colhead{Group}\\
\colhead{(1)} & \colhead{(2)} &	\colhead{(3)} 	&	\colhead{(4)} 	&	\colhead{(5)} 	&	\colhead{(6)} 	&	\colhead{(7)} 	&	\colhead{(8)} 	&	\colhead{(9)} 	&	\colhead{(10)} &	\colhead{(11)} 	&	\colhead{(12)} 	&	\colhead{(13)} 	&	\colhead{(14)} } 
\startdata
275	&	0.4018	&	4	&	0	\phm{00} &	0.0	\phm{0} &	-2.4	\phm{000} &	...	\phm{0} 	&	k	&	$<$8.00E-05	\phm{0}	&	SBab	&	...	&	71.1	&	-83.2	&	Core	\\																																																																																																																																																																																																																																		
276	&	0.4146	&	2	&	0	\phm{00} &	0.0	\phm{0} &	-0.9	\phm{000} &	...	\phm{0} 	&	k	&	$<$8.00E-05	\phm{0}	&	E?	&	...	&	-12.1	&	-80.5	&	Core	\\																																																																																																																																																																																																																																		
277	&	0.4052	&	4	&	0	\phm{00} &	6.3	\phm{0} &	3.9	\phm{000} &	6	\phm{0} 	&	a+k	&	$<$8.00E-05	\phm{0}	&	*	&	...	&	-10.7	&	-72.5	&	Core	\\																																																																																																																																																																																																																																		
278	&	0.4084	&	2	&	-15	\phm{00} &	2.2	\phm{0} &	-0.5	\phm{000} &	...	\phm{0} 	&	e(c)	&	3.05E-04	\phm{0}	&	*	&	...	&	-9.0	&	-54.5	&	Core	\\																																																																																																																																																																																																																																		
279	&	0.4099	&	4	&	0	\phm{00} &	0.0	\phm{0} &	-2.4	\phm{000} &	...	\phm{0} 	&	k	&	4.55E-04	\phm{0}	&	Sd	&	M?	&	98.3	&	-48.6	&	Core	\\																																																																																																																																																																																																																																		
280	&	0.4103	&	4	&	0	\phm{00} &	0.0	\phm{0} &	-2.4	\phm{000} &	...	\phm{0} 	&	k	&	$<$8.00E-05	\phm{0}	&	E/S0	&	...	&	-20.8	&	-25.8	&	Core	\\
\enddata
\tablecomments{ Col. (1) Object--- galaxy index number; Col. (2) z--- redshift; Col. (3) Q--- spectrum quality/redshift reliability (1=excellent 3=fair); Col. (4) \OII--- \OII rest frame equivalent width in angstroms; Col. (5) \Hd--- \Hd rest frame equivalent width in angstroms. A blank in columns 4 and/or 5 signifies an uncertain measurement; Col. (6) \DHd--- excess \Hd strength, as defined in \S5; Col. (7) $\sigma(\Hd)$--- the rest frame width of the \Hd absorption line, only measured for stronger lines; Col. (8) class--- spectral class, as defined by D99, and discussed in   \S5; Col. (9) F$_{24\micron}$--- \tfm flux measured by {\em Spitzer}, in Jansky's, including upper limits from D09; ''...'' indicates that the object is outside of the {\em Spitzer} field; Col. (10) morph--- morphological type ''...'' indicates that the object is outside of the {\em HST} fields, ''x'' = stellar;  Col. (11) int--- out best guess about the nature of distortions T = tidal tail, I = interaction with neighbor, M = merger, question marks denote uncertain cases;  Cols. (12,13) $\Delta \alpha, \Delta \delta$--- galaxy coordinate, {\em in arc seconds for both coordinates} relative to the cluster center as determined below; Col. (14) Group--- subgroup membership, as determined in   \S3.}
\end{deluxetable}

%
%
\begin{deluxetable}{rrccrrrrccc}
\tablewidth{0pt}
\tablenum{3}
\tabletypesize{\footnotesize}
\tablecaption{Field Galaxy Sample $0.30 \le z \le 0.55$}
\tablehead{\colhead{Cluster}		&	\colhead{Object} 	&	\colhead{z} 	&	\colhead{Q} 	&	\colhead{\OII} 		&	\colhead{\Hd} 		&	\colhead{\DHd} 		&	\colhead{$\sigma(\Hd)$} 		&	\colhead{class} 	&	\colhead{morph} 	&	\colhead{int} 	\\
\colhead{(1)} 		&	\colhead{(2)} 	&	\colhead{(3)} 	&	\colhead{(4)} 	&	\colhead{(5)} 		&	\colhead{(6)} 		&	\colhead{(7)} 		&	\colhead{(8)} 		&	\colhead{(9)} 	&	\colhead{(10)} 	&	\colhead{(12)} 	}
\startdata
3C295		&	33	\phm{00}	&	0.3286	&	2	&	-28	\phm{0}	&	3.6	\phm{0}	&	0.9	\phm{000}	&	...	\phm{0}	&	e(c)	&	...	&	...	\\
3C295		&	35	\phm{00}	&	0.3798	&	2	&	-22	\phm{0}	&	0.0	\phm{0}	&	-2.8	\phm{000}	&	...	\phm{0}	&	e(c)	&	S0/Sa	&	.	\\
A370		&	42	\phm{00}	&	0.3460	&	2	&	-8	\phm{0}	&	6.8	\phm{0}	&	4.3	\phm{000}	&	11	\phm{0}	&	e(a):	&	...	&	...	\\
A370		&	46	\phm{00}	&	0.3364	&	3	&	-57	\phm{0}	&	0.0	\phm{0}	&	-1.4	\phm{000}	&	...	\phm{0}	&	e(b)	&	...	&	...	\\
A370		&	47	\phm{00}	&	0.5485	&	2	&	-9	\phm{0}	&	...	\phm{0}	&	...	\phm{000}	&	...	\phm{0}	&	e(a)	&	...	&	...	\\
A370		&	48	\phm{00}	&	0.3286	&	3	&	-15	\phm{0}	&	0.0	\phm{0}	&	-2.7	\phm{000}	&	...	\phm{0}	&	e(c)	&	...	&	...	\\
A370		&	49	\phm{00}	&	0.4217	&	1	&	-10	\phm{0}	&	2.2	\phm{0}	&	-0.3	\phm{000}	&	...	\phm{0}	&	e(c)	&	...	&	...	\\
A370		&	50	\phm{00}	&	0.3474	&	2	&	-22	\phm{0}	&	...	\phm{0}	&	...	\phm{000}	&	...	\phm{0}	&	e(a)	&	...	&	...	\\
\enddata
\tablecomments{Col. (1) Cluster--- cluster field from which the galaxy was obtained; Col. (2) Galaxy--- galaxy index number, from Table 1 for A851, from D99 for other clusters; Col. (3) z--- redshift; Col. (4) Q--- spectrum quality/redshift reliability (1=excellent 3=fair); Col. (5) \OII--- \OII rest frame equivalent width in angstroms; Col. (6) \Hd--- \Hd rest frame equivalent width in \AA. A blank in columns 4 and/or 5 signifies an uncertain measurement; Col. (7) \DHd--- excess \Hd strength, as defined in \S5; Col. (8) $\sigma(\Hd)$--- the rest frame width of the \Hd absorption line, only measured for stronger lines; Col. (9) class--- spectral class, as defined by D99, and discussed in   \S5;  Col. (10) morph--- morphological type ''...'' indicates that the object is outside of the {\em HST} fields, ''x'' = stellar;  Col. (11) int--- out best guess about the nature of the distortion T = tidal tail, I = interaction with neighbor, M = merger, question marks denote uncertain cases}
\end{deluxetable}


\begin{deluxetable}{lcrr}
\tablewidth{0pt}
\tablenum{4}
\tablecaption{Subcluster Properties}
\tablehead{\colhead{Subcluster} & \colhead{$N_{gal}$} & \colhead{$<z>$} & \colhead{$\sigma(\kms)$}}
\startdata
Core & 80 & 0.4068 & 1079\phm{000}\\
N group & 11 & 0.3946 & 295\phm{000}\\
NW group & 10 & 0.4021 & 412\phm{000}\\
\enddata
\end{deluxetable}

%
%
\begin{deluxetable}{lccl}
\tablewidth{0pt}
\tablenum{5}
\tablecaption{Spectroscopic Classes}
\tablehead{\colhead{Class} & \colhead{\EWOII} & \colhead{\EWHd} & \colhead{Stellar Population}}
\startdata
k    & absent &  $< 3$\AA & passive \\
k+a  & absent & 3\AA\  -- 8\AA & weak post-starburst\\
a+k  & absent & $\geq 8$\AA & strong post-starburst\\
e(c) & -40\AA\  -- -5\AA & $< 4$\AA & continuous star formation\\
e(a) & -40\AA\ -- -5\AA &$\geq 4$\AA & ongoing starburst\\
e(b) & $< -40$\AA & any & ongoing starburst or low z star formation\\
e(n) &  ... & ... & AGN spectrum\\
\enddata
\end{deluxetable}

%
%
\begin{deluxetable}{lcccc}
\tablewidth{6in}
\tablenum{6}
\tablecaption{Galaxy Populations of Various Environments}
\tablehead{\colhead{} & \colhead{Cluster Core $R<75\arcsec$} & \colhead{Main Cluster} & \colhead{Infalling Groups} & \colhead{Field $0.3\le z \le 0.55$}}
\startdata
$N_{tot}$ & 29 & 80 & 21 &76\\
$f_{active}$ & 0.48 & 0.51 & 0.67 & 0.79\\
$f_{SB}/f_{active}$ & 0.71 & 0.44 & 0.71 & 0.16\\
$f_{PSB}/f_{SB}$ & 0.00 & 0.65 & 0.20 & 0.18\\
$f_M/f_{SB}$ & 0.78 & 0.54 & 0.00 & 0.00\\
\enddata
\end{deluxetable}

%
%

%
%
\begin{figure}
\includegraphics[scale=0.9,angle=0] {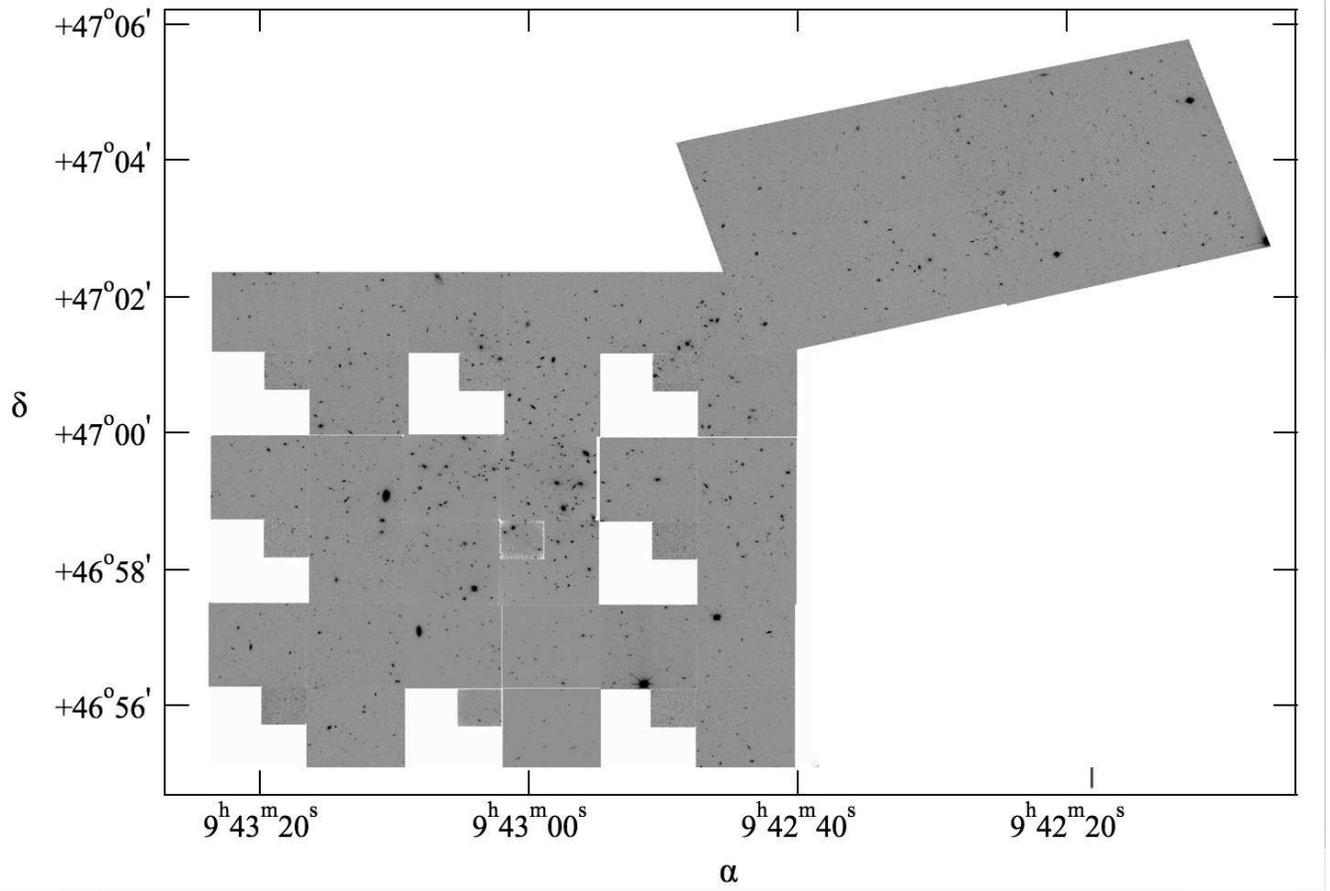}
\caption{A mosaic of all of our {\em HST} imaging in A851. This includes  one WFPC-1  from Cycle 1, two WFPC-2 fields from cycle 4, 7 WFPC-2 fields from Cycle 6, and 2 ACS fields from Cycle 16.}
\end{figure}

\clearpage
%
%
\begin{figure}
\includegraphics[scale=0.7,angle=270] {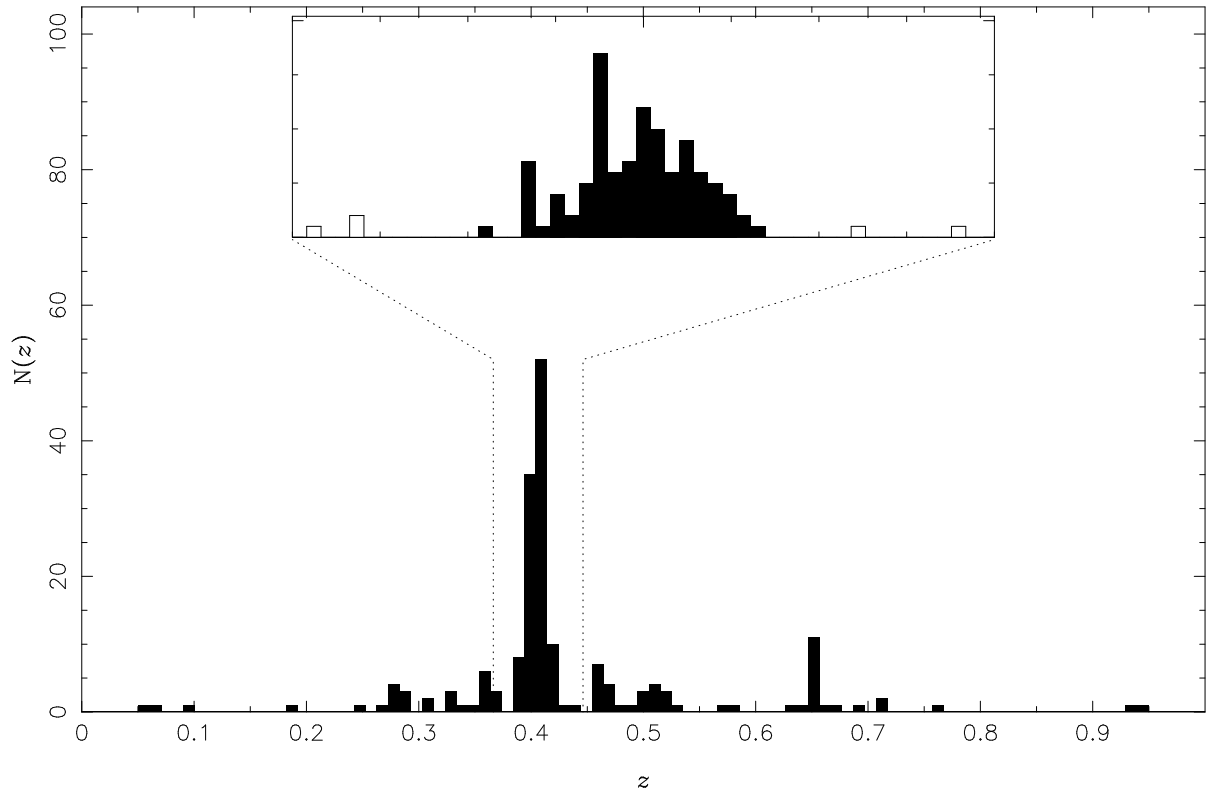}
\caption{The redshift distribution for cluster
members.  The two lowest-redshift galaxies in the distribution
have been eliminated as members of the surrounding supercluster.}
\end{figure}

%
%
\begin{figure}
\includegraphics[scale=0.75,angle=0] {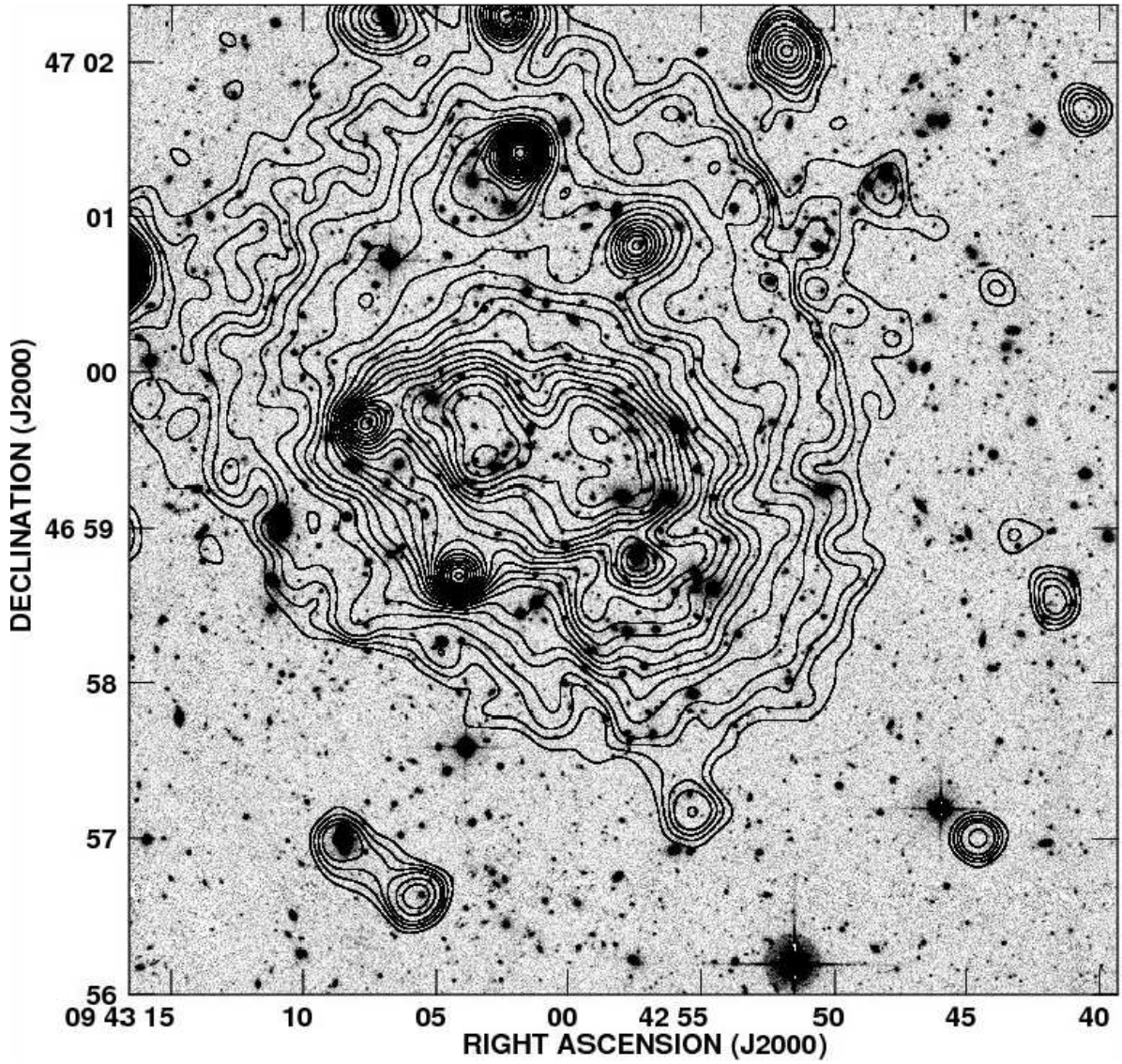}
\caption{The XMM X-ray image of the cluster Abell 851 overlaid on a  Megacam r band image. The double structure in the center shows a merger-in-progress, and there are signs of additional X-ray emission associated with infalling subgroups 1-2 Mpc from the cluster center, including the inner part of the Northwest filament/group.}
\end{figure}

%
%
\begin{figure}
\plottwo{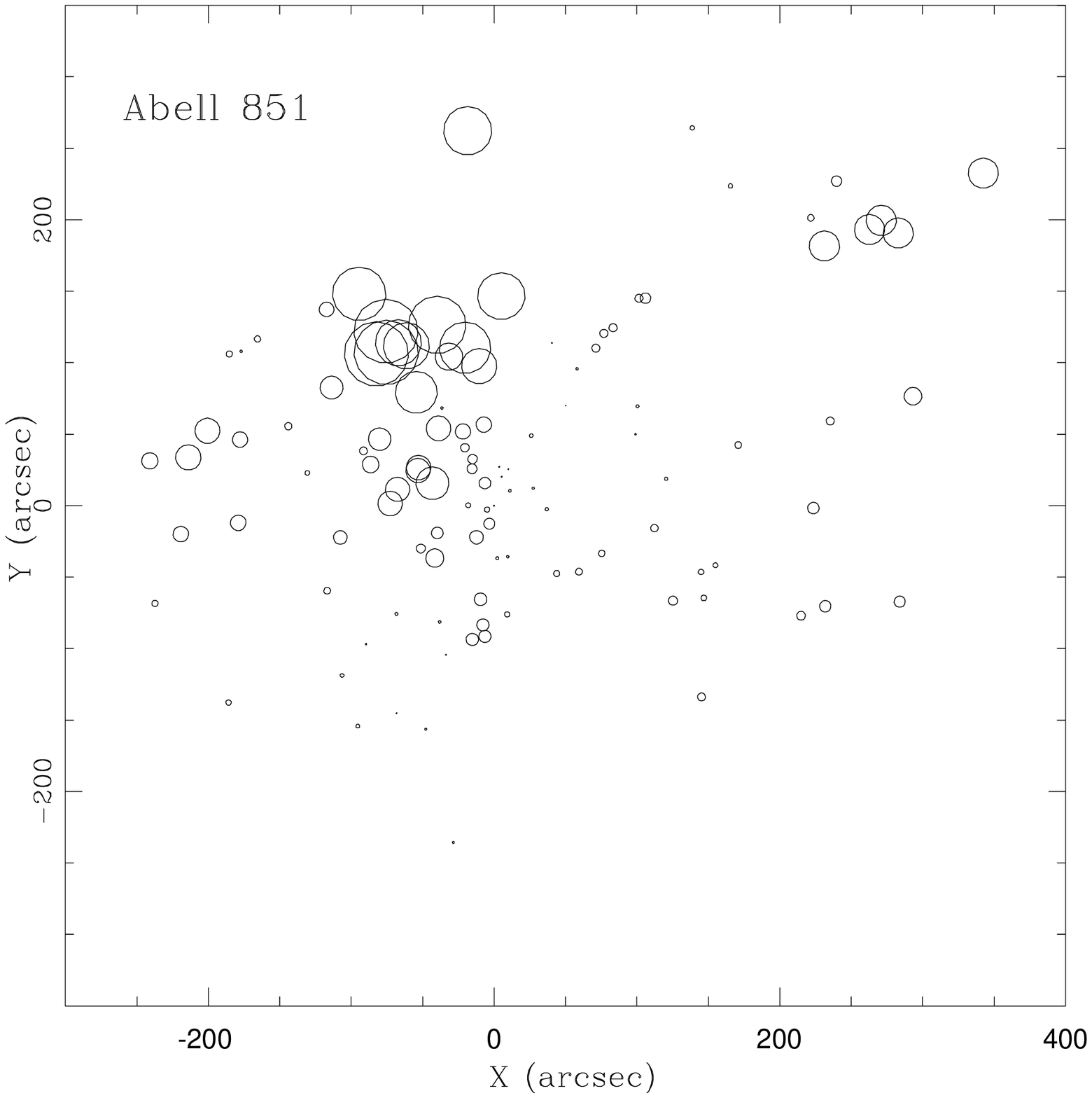}{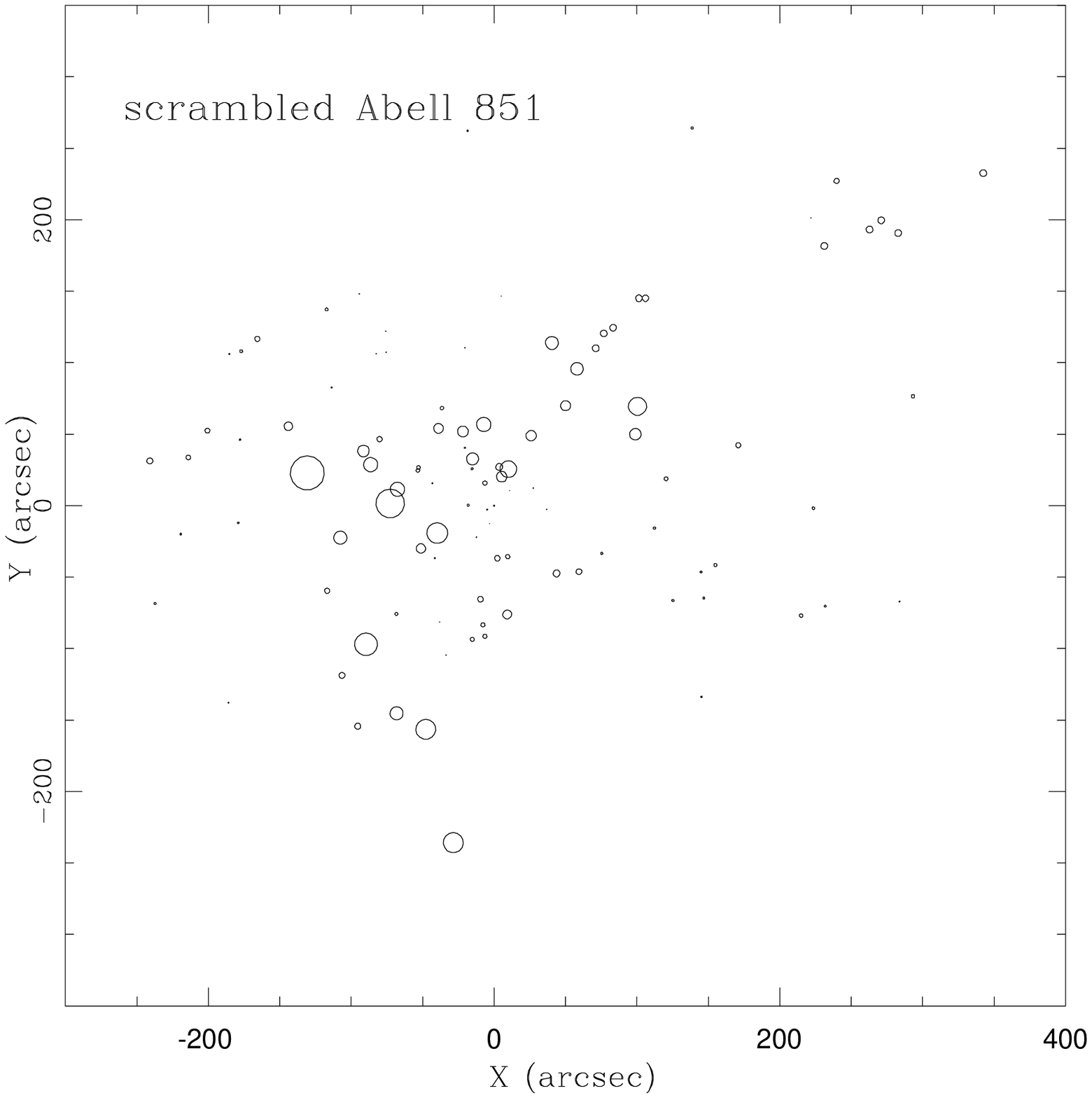}
\caption{left-- The Dressler-Shectman test for substructure.  The sizes of the circles at each galaxy location are proportional to the departure of the local velocity mean and dispersion from the global values. Large circles show significant departures
of local velocity and velocity dispersions from global means.  right-- A typical ``shuffled'' cluster, showing the size of expected deviations if the substructure in this cluster were not real.}
\end{figure}


\begin{figure}
\plotone{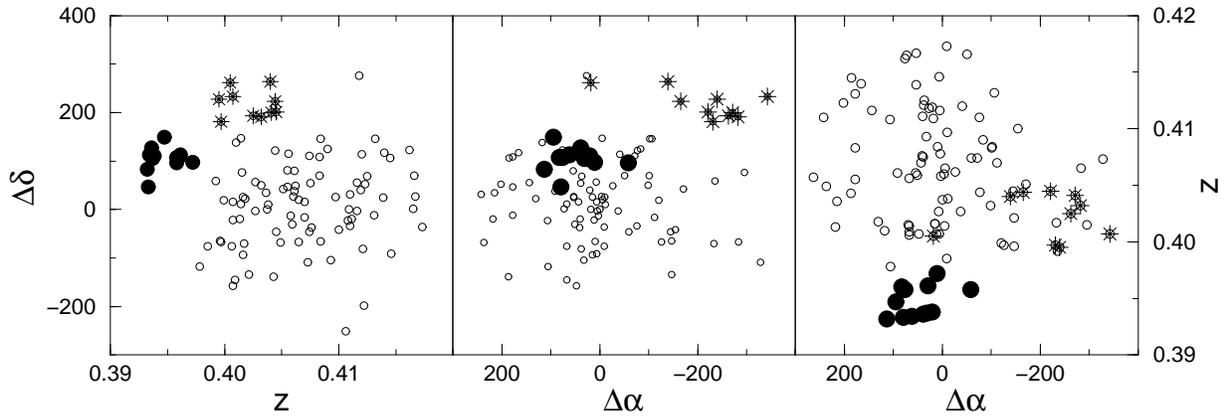}
\caption{The distribution of A851 cluster members in left-- redshift--declination space, center-- right ascension-declination space, and right-- right ascension--redshift space. Members of the North group are shown as filled circles and members of the Northwest group as stars.}
\end{figure}

%
%
\begin{figure}
\includegraphics[scale=0.27,angle=0,clip=true] {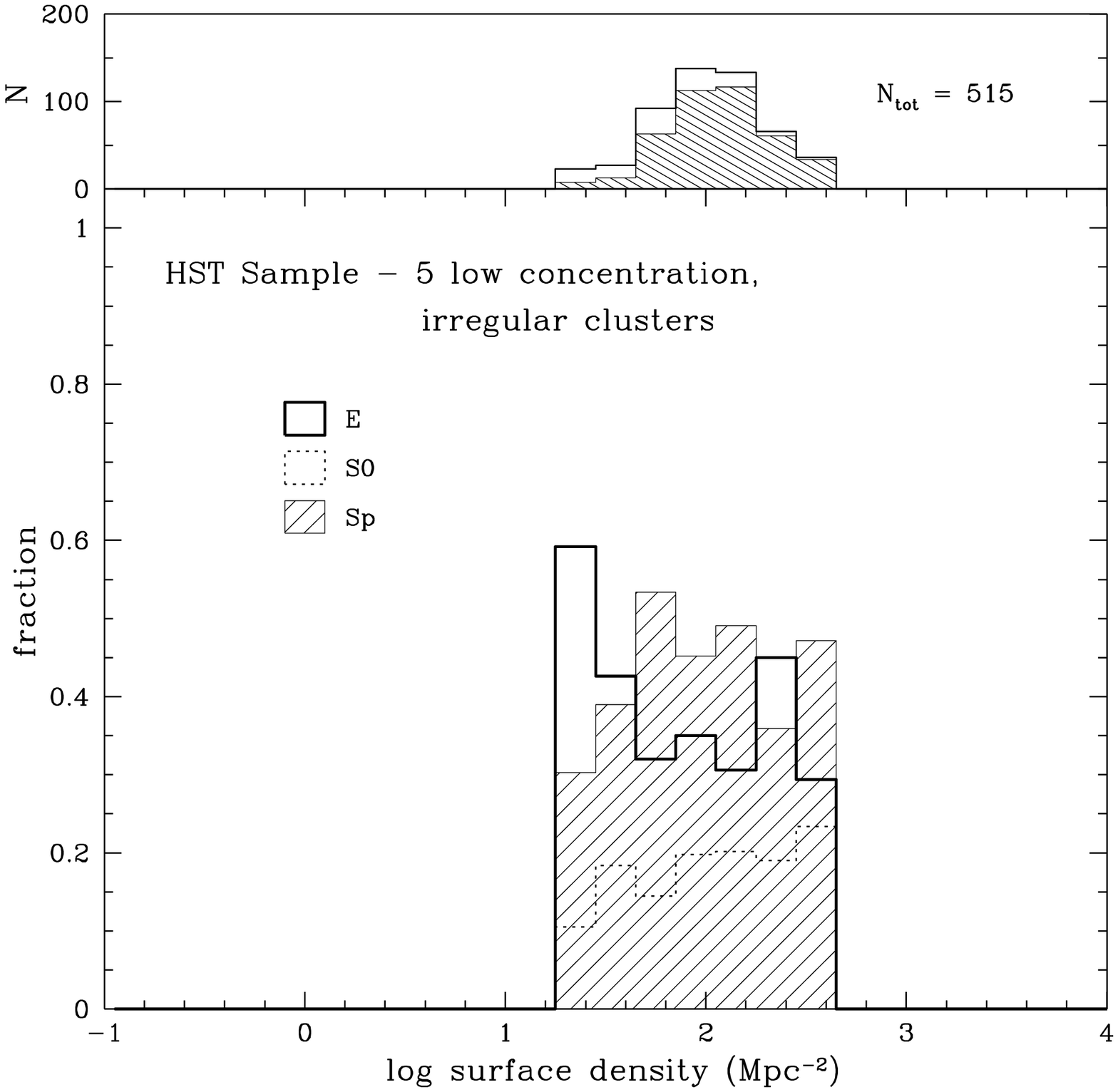}
\hfil
\includegraphics[scale=0.27,angle=0,clip=true] {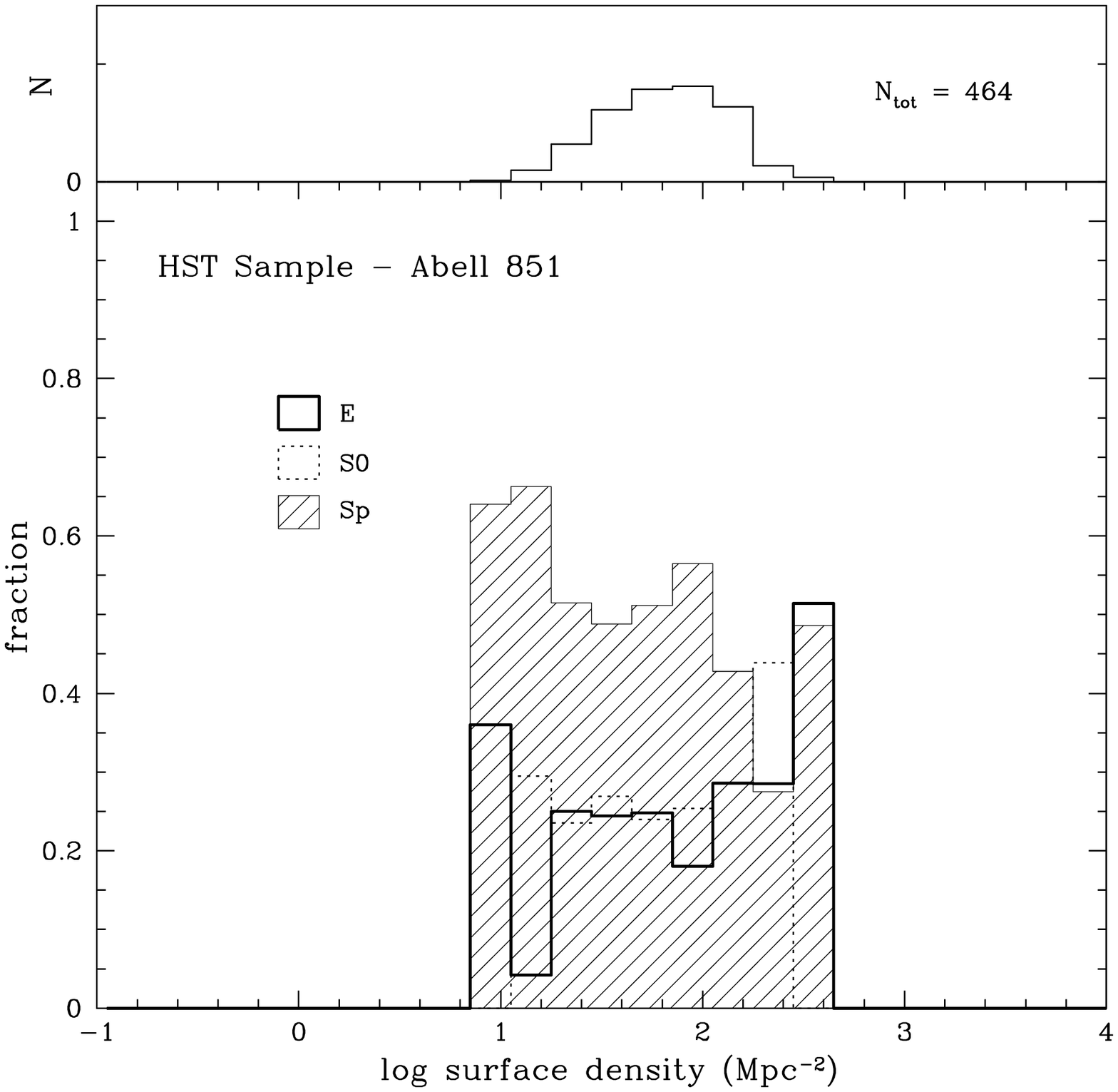}
\hfil
\includegraphics[scale=0.27,angle=0,clip=true] {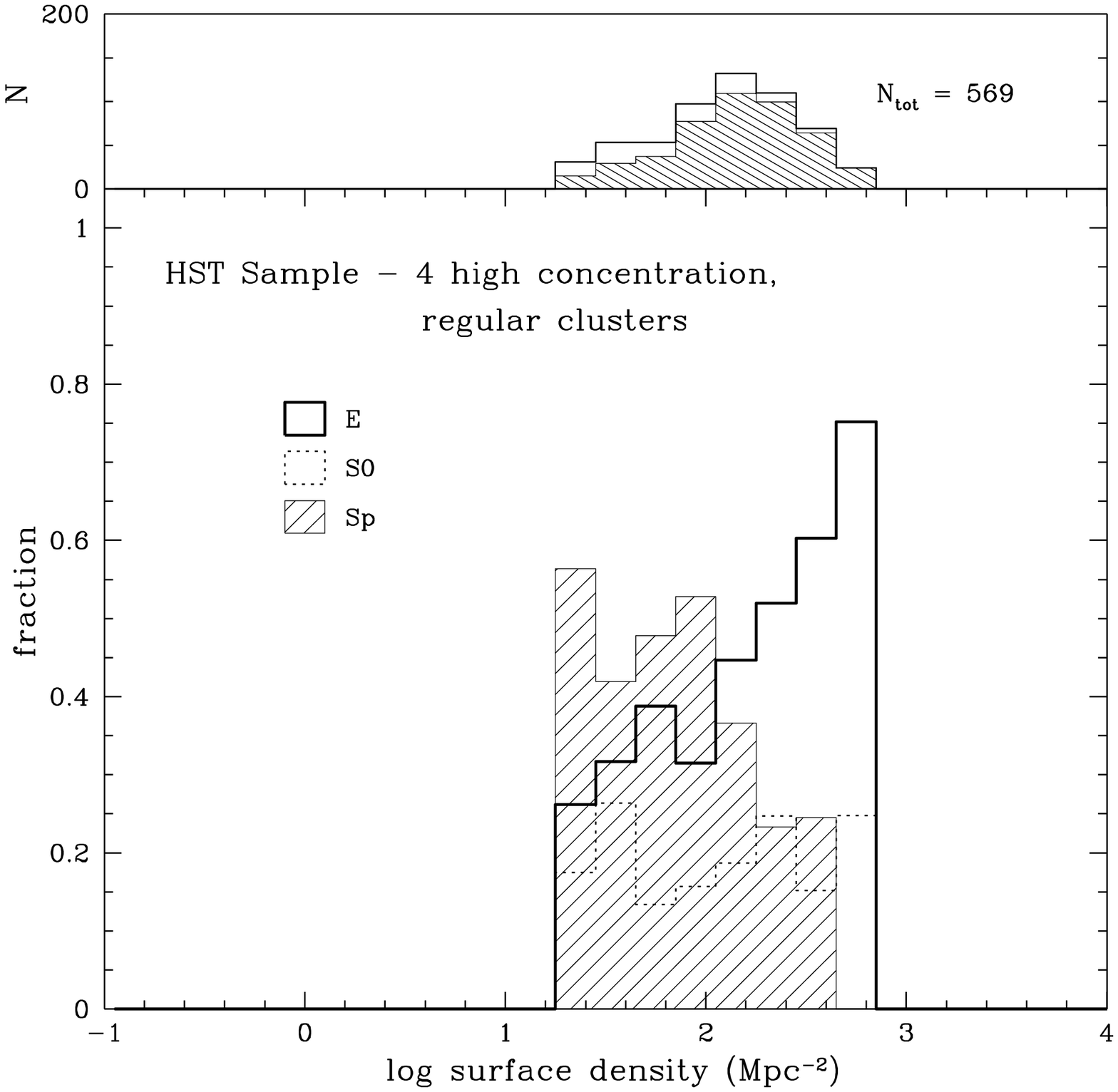}
\caption{ The morphology-density relation for the extended field of Abell 851 (center).  The total sample of
galaxies is raised to 464, of which 120 are subtracted as field galaxies.The strong trends with density comparable
to what D97 found  for regular, concentrated clusters (right) in contrast with the absence of a morphology-density
relation for irregular clusters (left).    This suggests that the absence of relation for the irregular cluster found by D97
might simply be a result of the strong effects of projection for the small fields in D97, and that both regular and
irregular clusters at $z \sim 0.5$ might have strong morphology-density relations as do all present-epoch clusters.}
\end{figure}

%
%
\begin{figure}
\includegraphics[scale=0.60,angle=0,clip=true] {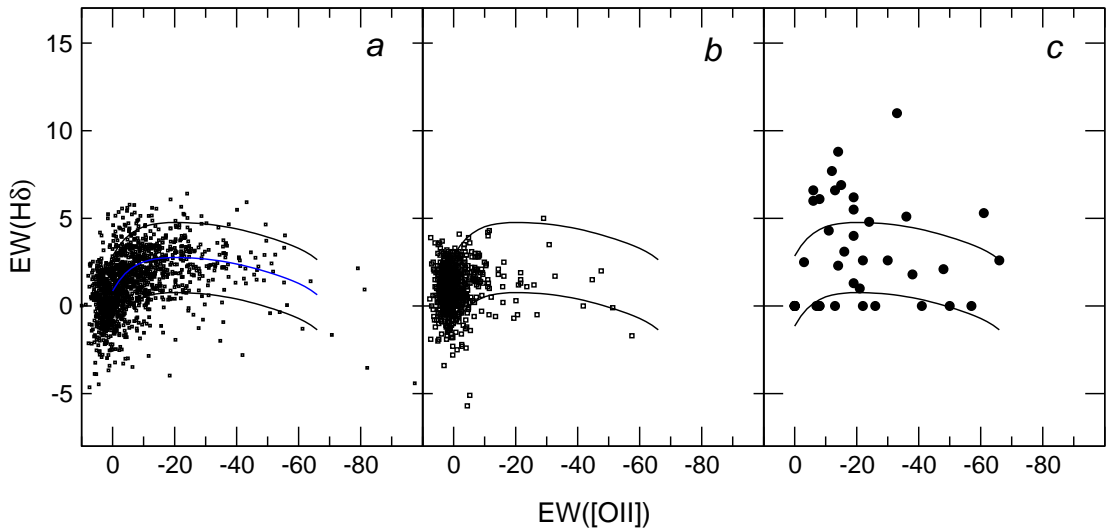}
\caption{Distribution of the equivalent widths of the \Hd\ and \OII lines in several low-redshift populations. In each plot the parallel lines enclose the expected locus of \Hd\ vs \OII in normal starforming galaxies. a) A sample of galaxies from the Las Campanas Redshift Survey, b) Nearby cluster galaxies, from the study by Dressler \& Shectman (1988a), c) A sample of merging galaxies, observed by Liu \& Kennicutt.}
\end{figure}

%

\begin{figure}
\epsscale{0.8}
\plotone{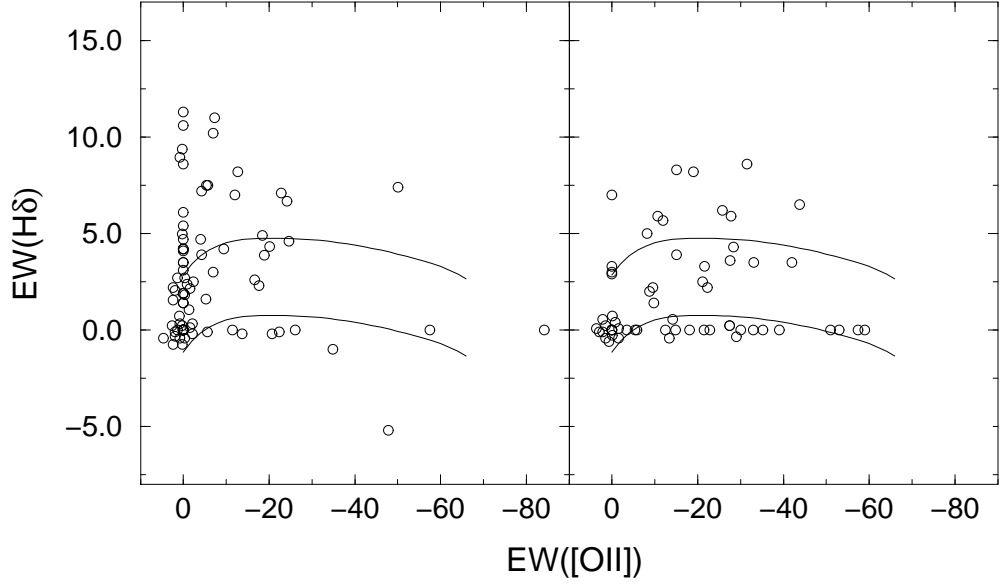}
\caption{Distribution of the equivalent widths of the \Hd\ and \OII lines in Abell 851 (a) and a sample of field galaxies with redshifts $ 0.30 \le z \le 0.55$ (b). As in fig. 9, the parallel lines enclose the locus of normal starforming galaxies.}
\end{figure}

%
%

\begin{figure}
\epsscale{0.35}
\plotone{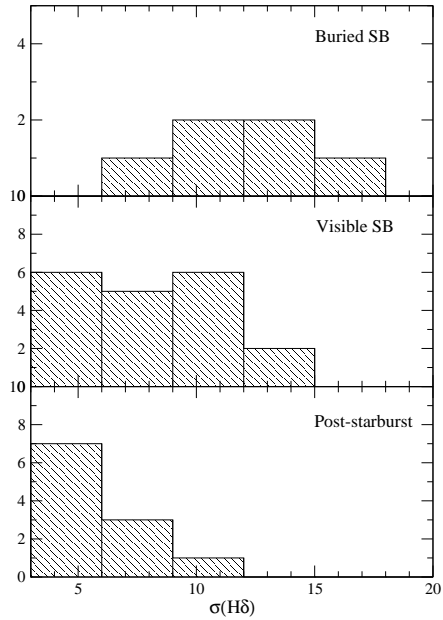}
\caption{Distribution of the widths of the \Hd\ line in top-- buried starbursts (k+a/a+k galaxies with \tfm emission), center-- visible starbursts (e(a) galaxies), and botton-- post-starbursts (k+a/a+k galaxies with no detected \tfm flux).}
\end{figure}


\begin{figure}
\epsscale{0.9}
\plotone{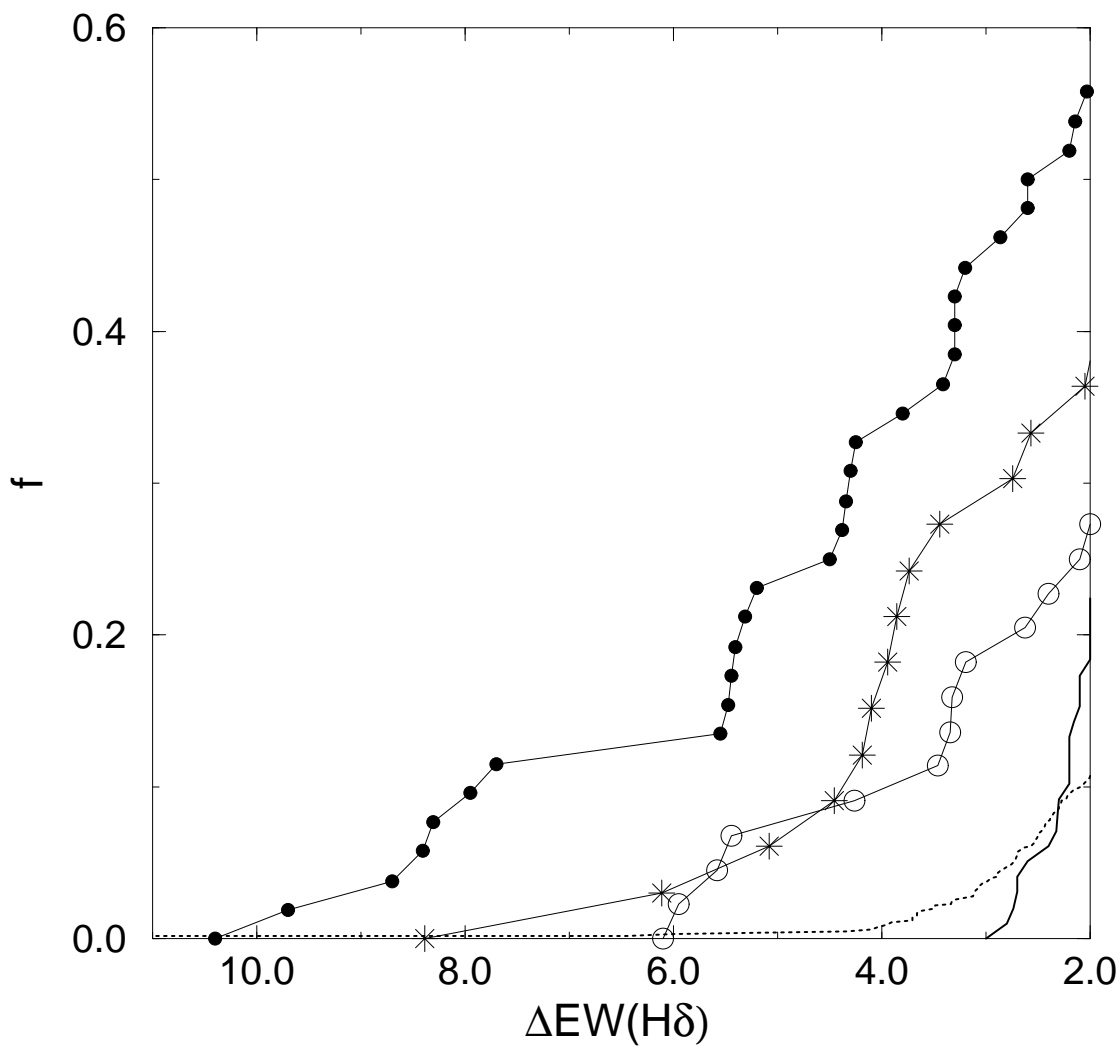}
\caption{Cumulative distribution of \Hd\ strengths
among ``active'' galaxies in various populations. Solid line--- LCRS
sample; dotted line--- DS cluster sample; stars--- LK merger sample;
filled circles--- A851; open circles--- intermediate z field sample.}
\end{figure}
\clearpage
%
%
\begin{figure}
\epsscale{1.0}
\plotone{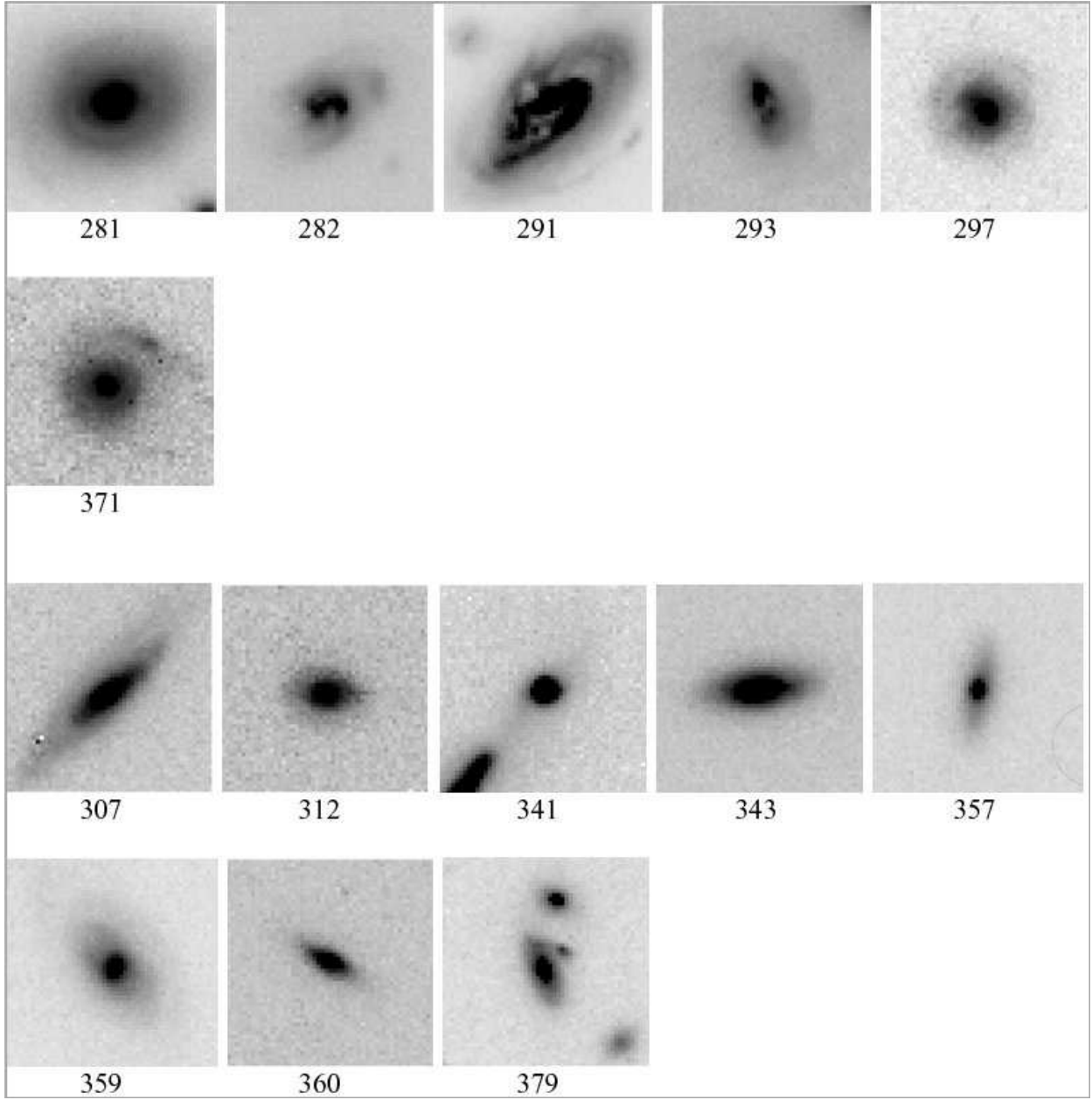}
\caption{Images of a+k and k+a type galaxies in A851. The top group consists of objects with detected \tfm flux, the bottom group consists of those without detectable \tfm flux. The top group all appear to be disturbed objects, probably mergers, while the lower group are mostly rather normal early-type disk galaxies.}
\end{figure}

%
%
\begin{figure}
\includegraphics[scale=0.9,angle=0] {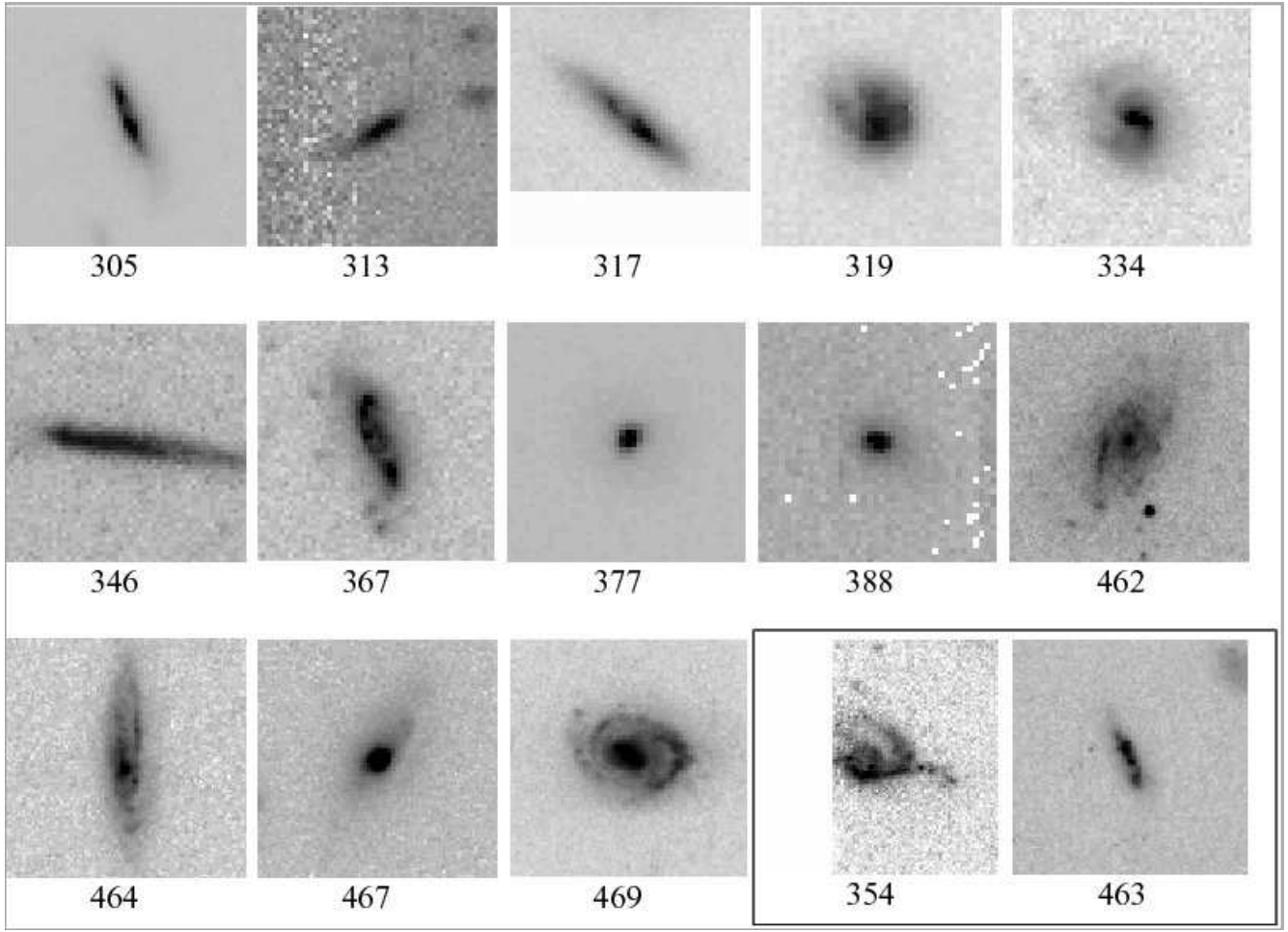}
\caption{Images of e(a) and e(b) type galaxies in A851. The last two, within a box, are e(b)'s; {\em Spitzer} observations confirm that object 354 is a starburst but object 463 is beyond the {\em Spitzer} area. The remainder of the objects are e(a)'s.}
\end{figure}

%
%
\begin{figure}
\includegraphics[scale=0.7,angle=0,clip=true] {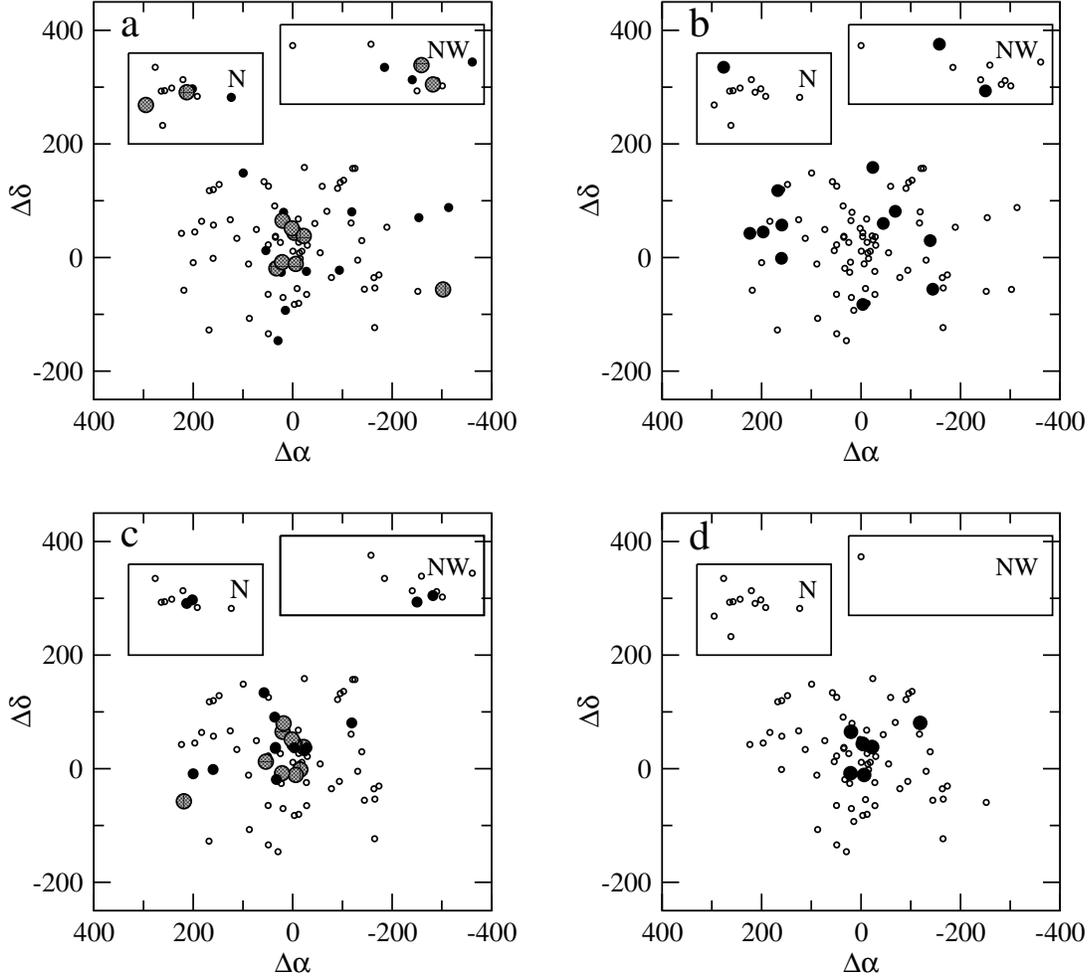}
\caption{(a) Distribution on the sky of starburst galaxies in A851. Open circles --- all cluster members; small black circles --- starbursts; large gray circles --- young starbursts with $\sigma(\Hd) > 11$\,\AA. (b) Distribution of post-starburst galaxies. Open circles --- all cluster members; filled circles --- post-starbursts. (c) Distribution of distorted galaxies. Open circles --- all cluster members with {\em HST} imaging; large filled circles --- M types, small filled circles  --- M?, T, T?, I, I?, and ? types. (d)  Distribution on the sky of buried, decaying starburst galaxies. Open circles --- all cluster members in the {\em Spitzer} survey area; filled circles --- galaxies with strong \Hd, detected \tfm flux, and no \OII.}
\end{figure}

%
%
\begin{figure}
\includegraphics[scale=0.7,angle=270,clip=true] {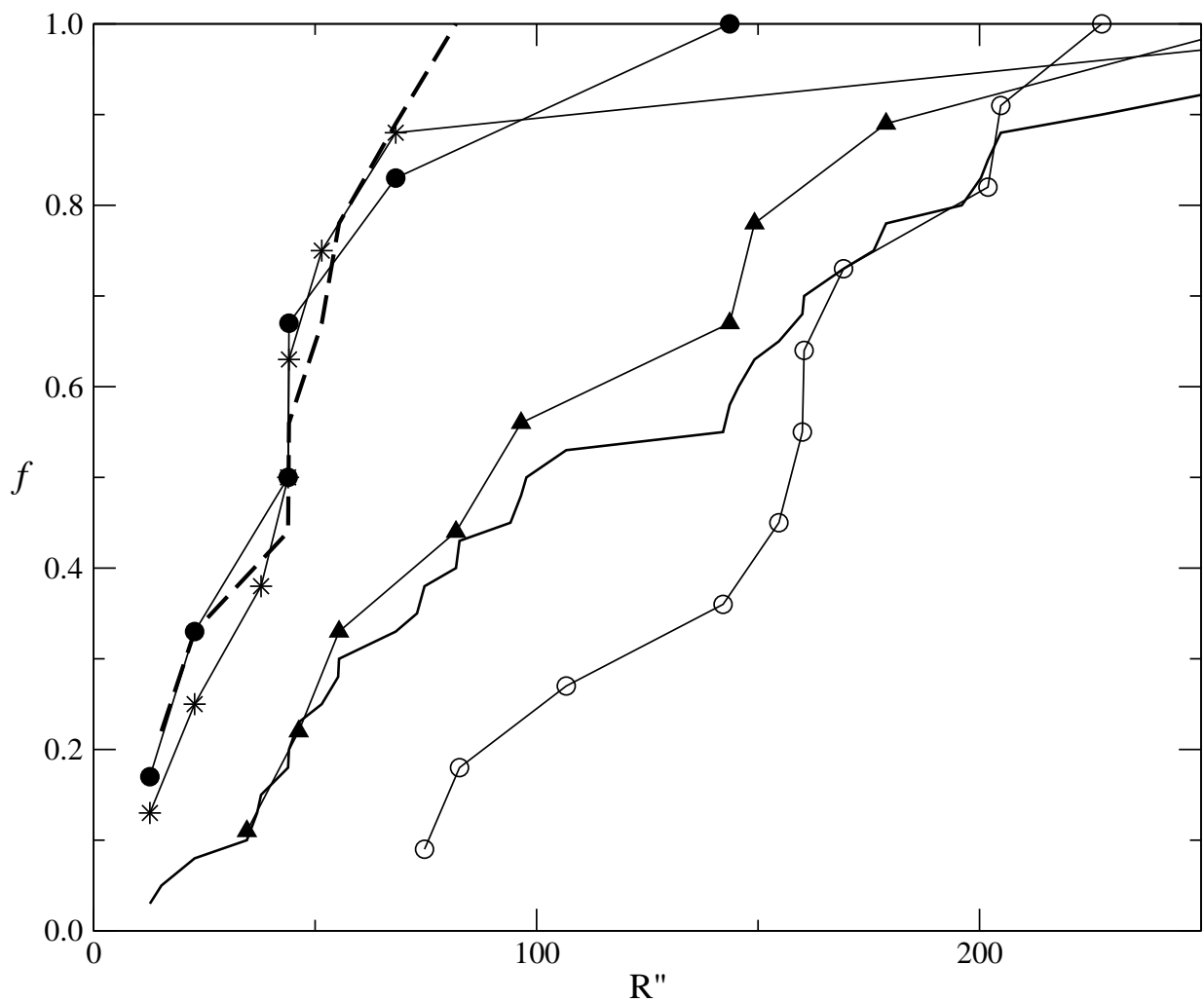}
\caption{Cumulative radial distribution of various galaxy subsets within the main cluster. Solid line --- all active galaxies, filled circles --- buried, decaying starbursts, stars- young starbursts $\sigma(\Hd) \ge 11$\,\AA, triangles --- old starbursts $\sigma(\Hd) < 11$\,\AA, dashed line --- M-class objects, open circles --- post-starbursts.}
\end{figure}

%
%
\begin{figure}
\plotone{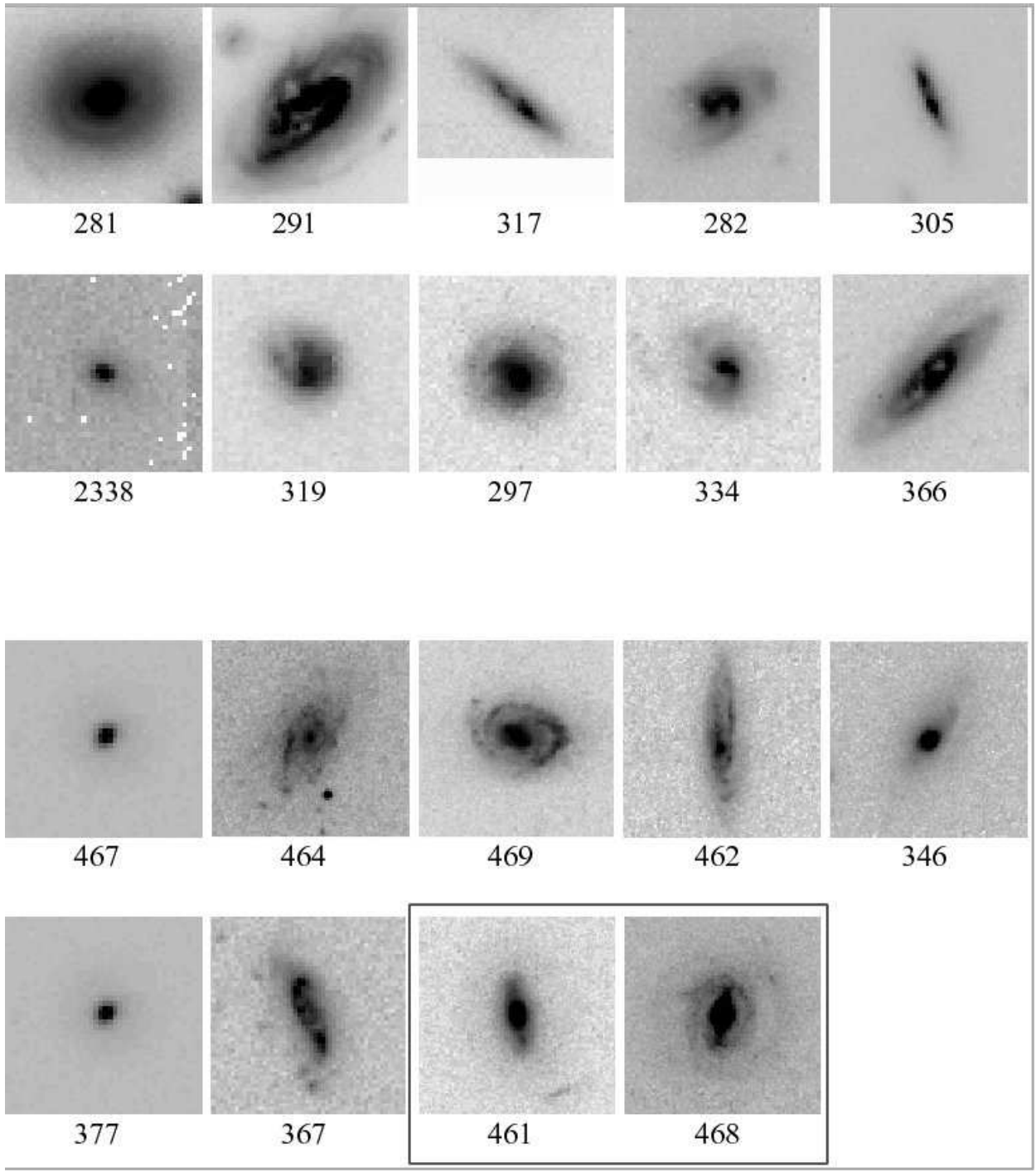}
\caption{Images of starburst galaxies within 75 arcsec of the cluster core (top group) and in the North and Northwest groups (bottom group). Objects within the box are a+k/k+a objects outside of the {\em Spitzer} survey area.}
\end{figure}

\end{document}